\documentclass[onecolumn,tighten,times]{aastex62}
\usepackage{txfonts}
\usepackage{newlfont}

\usepackage{amssymb}
\usepackage{wrapfig}
\usepackage{cancel}

\usepackage{relsize,exscale}
\makeatletter

\usepackage{array,multirow,graphicx}
\usepackage{dcolumn}
\usepackage{txfonts}
\usepackage{newlfont}
\usepackage{bm}
\usepackage[figtopcap]{subfigure}

\shorttitle{Anisotropic compact stars in the mimetic gravitational theory}
\shortauthors{G.~G.~L.~Nashed}

\begin{document}

\title{\uppercase{Anisotropic compact stars in the mimetic gravitational theory}}

\author{G.~G.~L.~Nashed}
\affiliation{Centre for theoretical physics, the British University in Egypt, 11837 - P.O. Box 43, Egypt}

\correspondingauthor{G.~G.~L.~Nashed}
\email{nashed@bue.edu.eg}


\begin{abstract}
In this paper, we consider  the mimetic gravitational theory to derive  a novel category of anisotropic star models. To end and to put the resulting differential equations into a closed system, the form of the metric potential $g_{rr}$ as  used by Tolman \citep{PhysRev.55.364} is assumed as well as a linear form of the equation-of-state. The resulting energy-momentum components, energy-density, and radial and tangential pressures contain five constants; three of these are determined through the junction condition, matching the interior with the exterior Schwarzschild solution the fourth  is constrained by the  vanishing of the radial pressure on the boundary and the fifth is constrained by a real compact star. The physical acceptability  of our model is tested using the data of the pulsar 4U 1820-30. The stability of this model is evaluated  using the Tolman-Oppenheimer-Volkoff equation and the adiabatic index and it is shown  to be  stable. Finally, our model is challenged with other compact stars demonstrating that it is consistent with those stars.
\end{abstract}



\section{Introduction}\label{S1}
Inflation has achieved remarkable success in solving some of the most important cosmic puzzles, such as the problem of flatness, the problem of the  horizon, and the problems of fine-tuning \citep{Guth:1980zm,Albrecht:1982wi,Linde:1981mu}. Furthermore, it has also been declared that if the pre-inflationary density fluctuations are allowed  to grow very largely over a short time ($\sim 10^{-33}$ sec.),  a clear and systematic explanation of the construction of the large-scale and anisotropic cosmic microwave background  structures is supplied \citep{Mukhanov:1981xt,Guth:1982ec,Hawking:1982cz,Starobinsky:1982ee}.

In Einstein's general relativity (GR), the scalar field which is coined as "inflaton" is responsible for executing inflation \citep{Mukhanov:1981xt}. In the GR realm, many studies  have been carried out that have confirmed the  consistency of the scalar  with observations  \citep{Hossain:2014coa,martin2014best,Martin:2013tda,Geng:2015fla,Huang:2015cke}.  However, GR still faces many challenges in confronting the accelerated expansion of our universe, among others, which is supported by observational data \citep{Riess:1998cb,Riess:2004nr,Perlmutter:1998np,Ade:2015xua}. Modified gravitational theories are considered as being reliable  in identifying cosmic problems such as the flat spin curves of spiral galaxies or the late cosmic acceleration, without  the need for dark matter and dark energy \citep{Capozziello:2011et,Nojiri:2010wj,Nashed:2006yw,Nojiri:2017ncd,DeFelice:2010aj}. Modified gravitational theories lead to scale-dependent interactions and  preserve the results of Einstein's theory at scales of the solar system \citep{Capozziello:2011et}.

Another modified gravitational  theory that has aimed to investigate dark matter is the mimetic
dark matter model that was built by Chamseddine and Mukhanov \citep{Chamseddine:2013kea}. They constructed a model whose concept was based on the isolation of the scalar degree-of-freedom of the metric. This model  was able to predict dark matter as well as  dark energy \citep{Chamseddine:2014vna} and
many cosmological solutions \citep{Chamseddine:2014vna} that   resolve singularities \citep{Chamseddine:2016uef,Chamseddine:2016ktu} and construct a
ghost-free massive gravity model \citep{Chamseddine:2018qym}.

 The construction of the mimetic theory can be investigated as a specific form of  conformal transformation in which the new metrics, as well as the old ones are degenerate. The use of non-singular conformal transformation produces an  increase in the  number of degrees-of-freedom, and thus;  the longitudinal sector of gravity turns out to be dynamic  \citep{Deruelle:2014zza,Domenech:2015tca,Firouzjahi:2018xob,Shen:2019nyp,Gorji:2019rlm}. Generally, conformal transformation means a relation between  ${\textit g_{\alpha \beta}}$,  which is considered the physical one,  to $ \bar{{\textit g}}_{\alpha \beta}$, which is the  auxiliary metric, and $\zeta$, which is the scalar field,  through the following transformation:
 \begin{equation} \label{trans1}
{\textit g}_{\alpha, \beta}=-\left(\bar{{\textit g}}_{\mu \nu}  \partial^\mu \zeta  \partial^\nu \zeta \right) \bar{{\textit g}}_{\alpha, \beta}\,.
\end{equation}
The above transformation gives  the following constraint \citep{Gorji:2019rlm}
\begin{equation} \label{trans2}
{\textit g}^{\alpha \beta}\partial_\alpha \zeta  \partial_\beta \zeta=-1\,.
\end{equation}
Therefore, $\partial_\alpha \zeta$ is a space such that we consider the positive sign and  a time--like such that we consider the negative  sign.
The famous mimetic  theory is that with a negative sign, and in this study, the positive sign is dealt with an extension to the mimetic gravitational theory (MGT).  Equation (\ref{trans1}) tells us that we cannot  write  $\bar{{\textit g}}_{\alpha \beta}$ in terms of   ${\textit g}_{\alpha \beta}$  since it is  non-invertible \citep{Deruelle:2014zza}. The extra degree of freedom linked to the transformation (\ref{trans1}) accounts for the longitudinal sector of gravity. Starting with  the Einstein-Hilbert action  that contains the physical metric $g_{\alpha \beta}$ and carries out  the transformation   (\ref{trans1})  ${\textit g}^{\alpha \beta}$  and $\zeta$ which are dynamic scalar fields
 \citep{Chaichian:2014qba}.

Generally,  when studying compact stars a spherically symmetric spacetime that has isotropy can be supposed. The isotropy and homogeneity of astrophysical compact stellar objects  cannot considered  as the general physical
characteristics of these stellar objects;  even though they could supply us with some solvable features. This is because the fluid pressure can be considered
to have two different components that  provide the anisotropy, $\Delta =p_t-p_r$. Such a case is called inhomogeneous  where  the source of the inhomogeneity results from the tangential, $p_t$, and the radial,  $p_r$,  pressures therefore, the distribution of matter is not in  isotropic form.  The idea of anisotropy was first considered by \citep{Ruderman:1972aj} and then by many researchers \citep{canuto1974equation,Bowers:1974tgi,herrera1997local}. Currently,  the source of anisotropy can be thought of as being due to different factors:\vspace{0.21cm}\\
$\bullet$ Superfluid 3A, \hspace{1cm}$\bullet$ a core of the region where the density is very high,\hspace{0.8cm} $\bullet$ a mixture of fluids of different types, \vspace{0.21cm}\\ $\bullet$ different condensate states,  \hspace{1cm} $\bullet$ relativistic particles in the compact stars  \hspace{1cm} $\bullet$ phase transition, \vspace{0.21cm}\\ $\bullet$ rotational motion,  \hspace{1cm} $\bullet$ and the existence  of the magnetic field, among others \citep{ivanov2002maximum,PhysRevD.82.044052,
Rahaman:2011cw,kalam2012anisotropic,
Deb:2015vda,Shee:2015kqa,ElHanafy:2015egm,Maurya:2016oml,maurya2018role,Deb:2016lvi}.  \vspace{0.21cm}\\ This  study aims to derive an interior spherically symmetric solution in the frame of MGT and to test if this solution represents a true compact star.

The structure of this research is as follows: In Section \ref{S2}, the cornerstone of MGT is given and its field equations are derived.  In Section \ref{S3}, the field equations of MGT  are applied to a spherically symmetric line element that has two unknown functions. To close  the system of differential equations, we assume the metric potential $g_{rr}$ to have the form given by  Tolman \citep{PhysRev.55.364}, and assume a linear form of the  equation-of-state (EoS) between the density and radial pressures. In Section \ref{S4},  the necessary requirements that any true compact star should satisfy are stated. In Section \ref{property}, the physical properties of our model are given, showing that the metric potential and the energy-momentum components are well defined in the center of the star.   Because  our model has five constants of integration, we made a junction condition with an exterior solution, a Schwarzschild solution, and  assumed that the radial pressure vanishes at the boundary of the star. From these constraints and  the use of the radial EoS  four  constants are determined leaving the fifth one to be  constrained by the comparison with the true star. In Section \ref{data}, the mass and radius of the compact star 4U 1820-30, are used showing that our model satisfies all the conditions  required for any true compact star.  In Section \ref{stability} the Tolman-Oppenheimer-Volkoff (TOV) equation and the adiabatic index of our model are derived and the model's willingness to be stable is shown. Also, in Section \ref{stability}, other compact stars are studied and  their relevant constants  are derived, with these results being summarized  in  Tables  (\ref{Table1}) and (\ref{Table2}).  In Section \ref{S5},  the results of the present study are summarized.
\section{Brief summary of the mimetic gravitational theory}\label{S2}


The term  ``mimetic dark matter" was proposed in the literature by Mukhanov and Chamseddine \citep{Chamseddine:2014vna} however, such a term was already  known in the theories  of papers written at earlier times  \citep{Lim:2010yk,Gao:2010gj,Capozziello:2010uv,Sebastiani:2016ras}. The Lagrangian of the mimetic theory in four-dimensions takes the following form:
\begin{equation} \label{act}
{\cal L}:=\frac{1}{2\chi} \int d^4x \sqrt{-g(\bar{g}_{\mu \nu},\zeta)}R(\bar{g}_{\mu \nu},\zeta)-\int d^4x \sqrt{-g(\bar{g}_{\mu \nu},\zeta)}  {\cal L_m}\,,
\end{equation}
where $\chi$ is the  gravitational constant, $\chi =\frac{8\pi G}{c^4}$, $G$ is the Newtonian constant, $c$ is the speed of light, $g(\bar{g}_{\mu \nu},\zeta)$ is the determinant of the metric tensor, $\zeta$ is the scalar field, $R$ is
the Ricci scalar, and ${\cal L_m}$ is the  Lagrangian of matter. Using Eq. (\ref{act}) the  field equations of the MGT  can be obtained as follows:
\begin{equation} \label{fe3}
 G_\mu{}^\nu-(G-T)\partial_\mu \zeta  \partial^\nu \zeta=\chi T_\mu{}^\nu\,, \qquad \qquad \nabla_\mu\left[(G-T)\partial^\mu \zeta\right]=0\,,
\end{equation}
where, $ G_{\mu \nu}$ and $T_{\mu \nu}$ are the Einstein and stress energy tensors
and $G$ and $T$ are their traces with $G = -R$. The stress-energy tensor, $T_\mu{}^\nu$, is the energy-momentum tensor
for fluids whose configuration has anisotropy and it is
represented by:
\begin{eqnarray}
&&T_\mu{}^\nu{}=(p_t+\rho)u_\mu u^\nu+p_t\delta_\mu{}^\nu+(p_r-p_t)\xi_\mu \xi^\nu\, ,
\end{eqnarray}
where $u_\mu$ is the time-like vector defined as $u^\mu=[1,0,0,0]$ and $\xi_\mu$ is the unit radial vector with its space-like property, defined by $\xi^\mu=[0,1,0,0]$ such that $u^\mu u_\mu=-1$ and $\xi^\mu\xi_\mu=1$.

  This  study aims to apply the field equations (\ref{fe3}) to a spherically symmetric spacetime.
\section{Interior solution in mimetic theory}\label{S3}
In this section,  the field equations of MGT, Eq. (\ref{fe3}), are to be applied to a spherically symmetric spacetime. For this purpose we use the following metric:
\begin{equation}\label{met}
 ds^{2}=w(r)dt^{2}-w_1(r)dr^{2}-r^2\left(d\theta^{2}+r^2\sin^2\theta d\phi^{2}\right)\,,
\end{equation}
where $w(r)$ and $w_1(r)$ are unknown functions. By applying the field equations (\ref{fe3}) to the spacetime (\ref{met}), we obtain the following non-linear differential equations:
\begin{eqnarray}\label{fe}
&&{\frac {w'_1  r+
 w_1{}^{2}-w_1 }{ w_1{}^{2}{r}^{2}}}=8\pi\rho\,,\nonumber\\
&&\frac {1}{2w_1{}^{3}{r}^{2}w^{2}}\Bigg(2\,w' r\,w\,w_1{}^{2}-2\,
 w_1{}^{3}w^{2}+2\,w^{2}
w_1{}^{2}+ \zeta'^{2}\Big[\pi(16\,{r}^{2} w^{2}w_1{}^{2}\rho -16\,{r}^{2}w^{2}w_1{}^{2}p_r-32\, {r}^{2} w^{2}w_1{}^{2}p_t)-
4\,w'_1 r w^{2}\nonumber\\
&&-4\,
 w^{2}w_1{}^{2}+4\,w^{2}w_1 +4\,  w' r\,w w_1 - {r}^{2}w' w'_1 w +2\, {r}^{2} w''w w_1-{r}^{2} w'^{2}w_1\Big]\Bigg)
=-8\pi\,p_r\,,\nonumber\\
&&\frac {2\, w'_1\,w^{2}-2\,w'ww_1 +rw'w'_1 w-2\,rw''\,w\,w_1 +r w'^{2}\,w_1 }{4r w^{2}w_1{}^{2}}
=-8\pi\,p_t\,.
\end{eqnarray}
where $w\equiv w(r)$, $w_1\equiv w_1(r)$,  $\zeta\equiv \zeta(r)$, and $'$ is the ordinary derivative, i.e.,  $w'\equiv \frac{dw}{dr}$.
The above differential system  consists of  three independent equations in six unknown  functions;  $w$, $w_1$, $\rho$, $p_r$, $p_t$, and $\zeta$.  Therefore, three extra conditions are required to solve the aforementioned system. One of these conditions is  to assume  the metric potential $g_{rr}$  to have the form:
\begin{equation}\label{sol1}
 w_1=1+s_1r^2+s_2r^4\,,
\end{equation}
where $s_1$ and $s_2$ are constants to be determined from the matching
conditions.
The metric potential (\ref{sol1}) was proposed by Tolman
(1939) to model realistic compact stellar objects. Interestingly,  the same metric component is used  to describe relativistic anisotropic
stellar objects with a prescribed linear EoS of the form:
\begin{equation}\label{rad1}
 p_r=s_3\rho+s_4\,,
\end{equation}
where $s_3$ and $s_4$ are constants. Finally, to satisfy Eq. (\ref{trans2}) we assume the scalar field in the form:
\begin{equation}\label{sc1}
 \zeta=\int\frac{1}{\sqrt{w_1}} dr\,.
\end{equation}
Using Eq. (\ref{sol1}) in the first equation of (\ref{fe}) we obtain the form of the density; then, using this form with Eq. (\ref{sc1}) in the remainder  of Eqs. (\ref{fe}) and  (\ref{rad1}) we obtain the form of $w$ as follows:
\begin{equation}\label{sol11}
 w=s_5\, \left( 1+s_1{r}^{2}+s_2{r}^{4} \right) ^{s_3}{e^{\displaystyle\frac{\,{r}^{2} \left( 16\,\pi\,s_2\,s_4\,{r}^{4}+24\,\pi\,s_1\,s_4\,{r}^{2}+3
\,s_2\,s_3\,{r}^{2}+3\,s_2{r}^{2}+6\,\pi\,s_3\,s_1+48\,s_4+6\,s_1
 \right) }{12}}}
\,.
\end{equation}
Consequently, the physical quantities are obtained as
\begin{eqnarray}\label{psol}
&&\rho( r) =\frac {s_2{}^{2}{r}^{6}+2\,s_1s_2{r}^{4}+
 \left( 5\,s_2+s_1{}^{2} \right){r}^{2}+3\,s_1}{8\pi\, \left( 1+s_1{r}^{2}+s_2{r
}^{4} \right) ^{2}}\,,\nonumber\\
&& p_r (r) =\frac {1}{8\pi\, \left( 1+s_1{r}^{2}+s_2{r}^{4} \right)
^{2}}\Big[8\,\pi\,s_2^{2}\,s_4\,{r}^{8}+
 \left( s_2^{2}\,s_3+16\,\pi\,s_1\,s_2\,s_4 \right) {r}^{6}+ \left( 8\,
\,\pi\,s_1^{2}\, s_4+16\,\pi\,s_2\,s_4+2\,s_1\,s_2\,s_3 \right) {r}^{4}\nonumber\\
&&+
 \left( 5\,s_2\,s_3+s_1{}^2\,s_3+16\,\pi\,s_1\,s_4 \right) {r}^{2}
+3\,s_1\,s_3+8\,\pi\,s_4\Big]\,,\nonumber\\
&&
\end{eqnarray}
where we give the form of the tangential pressure and the anisotropic force in Appendix A.
It is important to mention that the anisotropic force is defined as $\frac{2\Delta}{r}$ and it is attractive if $p_r-p_t>0$ and will be repulsive
 if $p_r-p_t<0$. The mass contained within a radius $r$ of the sphere
is defined as:
\begin{equation}\label{mas}
m(r)={\int_0}^r \rho(\eta)\eta^2 d\eta\,.\end{equation}
Using Eq. (\ref{psol}) in Eq. (\ref{mas}),  we get
\begin{equation}\label{mas1}
m(r)=\frac {{r}^{3} \left(s_1+s_2{r}^{2} \right) }{16\pi \, \left( 1+s_1{r}
^{2}+s_2{r}^{4} \right) }\,.\end{equation}
The compactness parameter of a spherically symmetric
source with radius $r$  takes the form \citep{Singh:2019ykp}
\begin{eqnarray}\label{gm1}
&&u(r)=\frac {{r}^{2} \left(s_1+s_2{r}^{2} \right) }{16\pi \, \left( 1+s_1{r}
^{2}+s_2{r}^{4} \right) }\,.
\end{eqnarray}
In the next section, we will state the physical requirements that any viable stellar structure  must satisfy and ascertain whether the model  (\ref{psol})  with the form of tangential pressure given in Appendix A satisfies them.
\section{Requirements for a physically consistent stellar model}\label{S4}
Any  physical viable stellar model must  satisfy the following
conditions throughout the stellar configurations:\vspace{0.1cm}\\
$\diamond$ The gravitational metric potentials, $w(r )$ and $w_1(r )$,  and the components of the energy-momentum tensor $\rho$, $p_r$ , $p_t$  must  be well-defined at the center of the star, as well as being regular and  free from singularity  throughout
the interior of the star.\vspace{0.1cm}\\
$\diamond$  The density, $\rho$, must  be positive in
the interior of the  stellar model, i.e., $\rho\geq 0$. The value of the density at the center
of the star should be positive, finite and monotonically
decreasing toward  the boundary.\vspace{0.1cm}\\
$\diamond$  The radial and the tangential pressures
must be positive inside the configuration of the fluid  i.e.,
$p_r\geq0$, $p_t\geq0$. Also, the derivative of the density and the pressures must be
negative, i.e., $\frac{d\rho}{dr}< 0$, $\frac{dp_r}{dr}< 0$ and $\frac{dp_t}{dr}< 0$.
The radial pressure, $p_r$, must vanish at the boundary of the stellar model $r= \mathcal{R}$,
however, the tangential pressure, $p_t$, need not be zero at the boundary. Finally, at the center of the star the pressures should be equal meaning
 that the anisotropy  vanishes, i.e., $\Delta(r = 0) = 0$.\vspace{0.1cm}\\
$\diamond$ Any anisotropic fluid sphere must   fulfill  the following energy condition inequalities: \vspace{0.1cm}\\
(i) Null energy condition (NEC): $p_t+\rho > 0$, $\rho> 0$.\vspace{0.1cm}\\
(ii) Strong energy condition (SEC): $p_r+\rho > 0$, $p_t+\rho > 0$, $\rho-p_r-2p_t > 0$.\vspace{0.1cm}\\
(iii) Weak energy condition (WEC): $p_r+\rho > 0$, $\rho> 0$.\vspace{0.1cm}\\
(iv) Dominant energy condition (DEC): $\rho\geq | p_r|$ and
$\rho\geq | p_t|$ .\vspace{0.1cm}\\
$\diamond$  The interior metric potentials must join smoothly
with the Schwarzschild  exterior metric at the boundary.\vspace{0.1cm}\\
$\diamond$  For a stable configuration, the adiabatic index must be
greater than $\frac{4}{3}$.\vspace{0.1cm}\\
$\diamond$ According to Herrera,  the stability of the
anisotropic stars should satisfy $0>v_r{}^2-v_t{}^2>-1$ where $v_r$ and $v_t$ are
the radial and transverse speeds respectively \citep{HERRERA1992206}.\vspace{0.1cm}\\
$\diamond$ To obtain a  realistic model,  the causality condition must be satisfied  meaning that the speed of sound must be less than 1
(providing the speed of light $c = 1$) in the interior of
the star, i.e., $1\geq\frac{dp_r}{dr}\geq 0$, $1\geq \frac{dp_t}{dr}\geq 0$.\vspace{0.1cm}\\

We are now ready  to analyze  the aforementioned  necessary physical conditions to  test whether our model satisfies them.
\section{Physical properties  of the model}\label{property}
Let us  test the model (\ref{psol}) and ({\bf {\color {blue} A}}),  see if it is consistent  with realistic  stellar structures. For this aim,  the following topics need to be discussed:
\subsection{The non-singular model}
i- The metric potentials of this model satisfy:
\begin{equation}\label{sing}
w(0)=s_5\,\qquad  \textrm{and} \qquad w_1(0)=1\,.
\end{equation}
Equation (\ref{sing})  implies that the gravitational metric potentials are finite at the center of the configuration of the stellar model. Furthermore,  the derivatives of these potentials are finite at the center, i.e., $w'_{r=0}=w'_1{_{r=0}}=0$. The aforementioned conditions ensure that the metric is regular at the center and behaves well   throughout the interior of the stellar model.\vspace{0.1cm}\\
ii-Density, radial, and tangential pressures; at the center, have the form:
\begin{equation}\label{reg}
\rho(0)=\frac{3s_1}{8\pi}\,, \qquad  \qquad p_r(0)=p_t(0)=\frac{3s_1s_3+8\pi s_4}{8\pi}.
\end{equation}
Eq. (\ref{reg}) shows that the density is always positive if $s_1>0$,  and the anisotropy is vanishing at the center. The radial and tangential pressures  have positive values if  $s_4+\frac{3s_1s_3}{8\pi}>0$; otherwise, they become negative. Furthermore,  the Zeldovich condition \citep{1971reas.book.....Z} states that the radial pressure must be less than or equal to the density at the center i.e., $\frac{p_r(0)}{\rho(0)}\leq 1$. Using the Zeldovich condition in Eq. (\ref{reg}), we obtain:
\begin{equation}\label{reg1}
s_1\leq \frac{3s_1s_3+8\pi s_4}{3}.
\end{equation}
iii-The derivative of the energy density, radial, and tangential
pressures  of the  model are, respectively:
\begin{eqnarray}\label{dsol}
&&\rho'=-\frac {r \left( 12\,{s_2}^{2}{r}^{4}+3\,{r}^{6}{s_2}^{2}s_1+{r}^{8}{s_2
}^{3}+13\,s_1{r}^{2}s_2+3\,{r}^{4}s_2{s_1}^{2}-5\,s_2+5\,{s_1}^{2}+{r}^{2}{s_1}^{3}
 \right) }{4\pi\, \left( 1+s_1{r}^{2}+s_2{r}^{4} \right) ^{3}}\,, \nonumber\\
 && p'_r=s_3\rho'\,,
\end{eqnarray}
where $\rho'=\frac{d\rho}{dr}$, $p'_r=\frac{dp_r}{dr}$ and the form of $p'_t=\frac{dp_t}{dr}$ is given in Appendix B. Eqs. (\ref{dsol}) and
({\bf {\color {blue} B}}) show that the gradients of the density, radial, and tangential pressures are negative as  will be shown in their plotting .\vspace{0.1cm}\\
iv-The radial and tangential velocities of sound (c = 1) are
obtained as:
\begin{eqnarray}\label{dso2}
&&v_r{}^2=\frac{dp_r}{d\rho}=s_3\,,
\end{eqnarray}
where the form of tangential speed is given in Appendix C.
\subsection{Matching conditions}
We assume that the exterior spacetime of a non-rotating star is empty and is described by the exterior Schwarzschild solution that is given by the following form:
\begin{eqnarray}\label{Eq1}  ds^2= -\left(1-\frac{2M}{r}\right)dt^2+\left(1-\frac{2M}{r}\right)^{-1}dr^2+r^2d\Omega^2,
 \end{eqnarray}
where $M$ is the total mass  and $r>2M$.
It is necessary to match the interior spacetime metric  (\ref{sol1}) and (\ref{sol11}) with
the exterior Schwarzschild spacetime metric  (\ref{Eq1}) at the
boundary of the star where $r = \mathcal{R}$. The continuity of the metric
functions across the boundary $r = \mathcal{R}$ gives the conditions
\begin{eqnarray}\label{Eq2}
w(r=\mathcal{R})=\left(1-\frac{2M}{\mathcal{R}}\right), \qquad \qquad w_1(r=\mathcal{R})=\left(1-\frac{2M}{\mathcal{R}}\right)^{-1}.
 \end{eqnarray}
Furthermore, we use the fact that the radial pressure approaches zero at a finite value of the radial parameter  $r$ r that coincides with the radius of the star  $\mathcal{R}$. Therefore, the radius of the star can be obtained using the physical condition $p_r (r =\mathcal{R}) = 0$. From the above conditions, we obtain the constraints on the constants $s_1$, $s_2$ and $s_5$. Using the above constraints, we obtain the constants  $s_1$, $s_2$ and $s_5$ as follows:
\begin{eqnarray}\label{Eq3}
&&s_1=-\frac {8\,s_3\,{M}^{2}-5\,s_3\,\mathcal{R}M-4\,s_4\,\pi\,{\mathcal{R}}^{4}}{
 {\mathcal{R}}^{2}s_3\left( \mathcal{R}-2\,M \right) ^{2}}\,, \qquad s_2=\frac {-3\,s_3\,\mathcal{R}M+4\,s_3\,{M}^{2}-4\,s_4\,\pi\,{\mathcal{R}}^{4}}{{\mathcal{R}}^
{4}s_3\, \left(\mathcal{R}-2\,M \right) ^{2} }\,,\nonumber\\
&&s_5=\frac{\left( \mathcal{R}-2\,M \right)}{\mathcal{R}}e^{\frac{
 \left( 36\,{M}^{2}-21\,\mathcal{R}M \right) {s_3}^{2}-12\, \ln  \left( {\frac {\mathcal{R}}{\mathcal{R}-2\,M}} \right)^{\left( \mathcal{R}-2M \right) ^{2}{s_3}^{2}} + \left( 36\,{M}^{2}-64
\,{M}^{2}{\mathcal{R}}^{2}s_4\,\pi+120\,Ms_4\,\pi\,{\mathcal{R}}^{3}-21\,\mathcal{R}M-60\,s_4
\,\pi\,{\mathcal{R}}^{4} \right) s_3-32\,{\mathcal{R}}^{6}{s_4}^{2}{\pi}^{2}-12\,
s_4\,\pi\,{\mathcal{R}}^{4} }{12 \left( \mathcal{R}-2\,M \right)^{2}{s_3}}}
\,.\nonumber\\
&&
 \end{eqnarray}
  The  constant $s_4$,  is determined from the fact that $p_r(r=\mathcal{R})=0\Rightarrow s_4=-s_3\rho(r=\mathcal{R})$ and the constant $s_3$ remains arbitrary so that its value will be adjusted with the real compact star.
 \section{Matching the model with the realistic compact stars} \label{data}
 Let us now consider the previous physical conditions of the model derived to test it using the masses and radii of the observed pulsars. To support our model, the pulsar \textrm{4U 1820-30} whose estimated mass and
radius are $M = 1.46\pm 0.21 M_\circledcirc$ and $\mathcal{R} =11.1\pm1.8$ km, respectively, will be studied \citep{Gangopadhyay:2013gha,Das:2021qaq,Roupas:2020mvs}.
The maximal values  $M=1.67 M_\circledcirc$ and  $\mathcal{R}=12.9$km can be used as the  input parameters.
The boundary conditions are adopted  to determine the
constants  $s_1=0.01294823993$, $s_2= -0.00005537504986$, and $s_5= 0.5917241379\,e^{- 0.7529881502- 1.480957408\,s_3}$.

\begin{figure}
\centering
\subfigure[~Metric potentials  (\ref{sol1}) and (\ref{sol11})]{\label{fig:pot1}\includegraphics[scale=0.3]{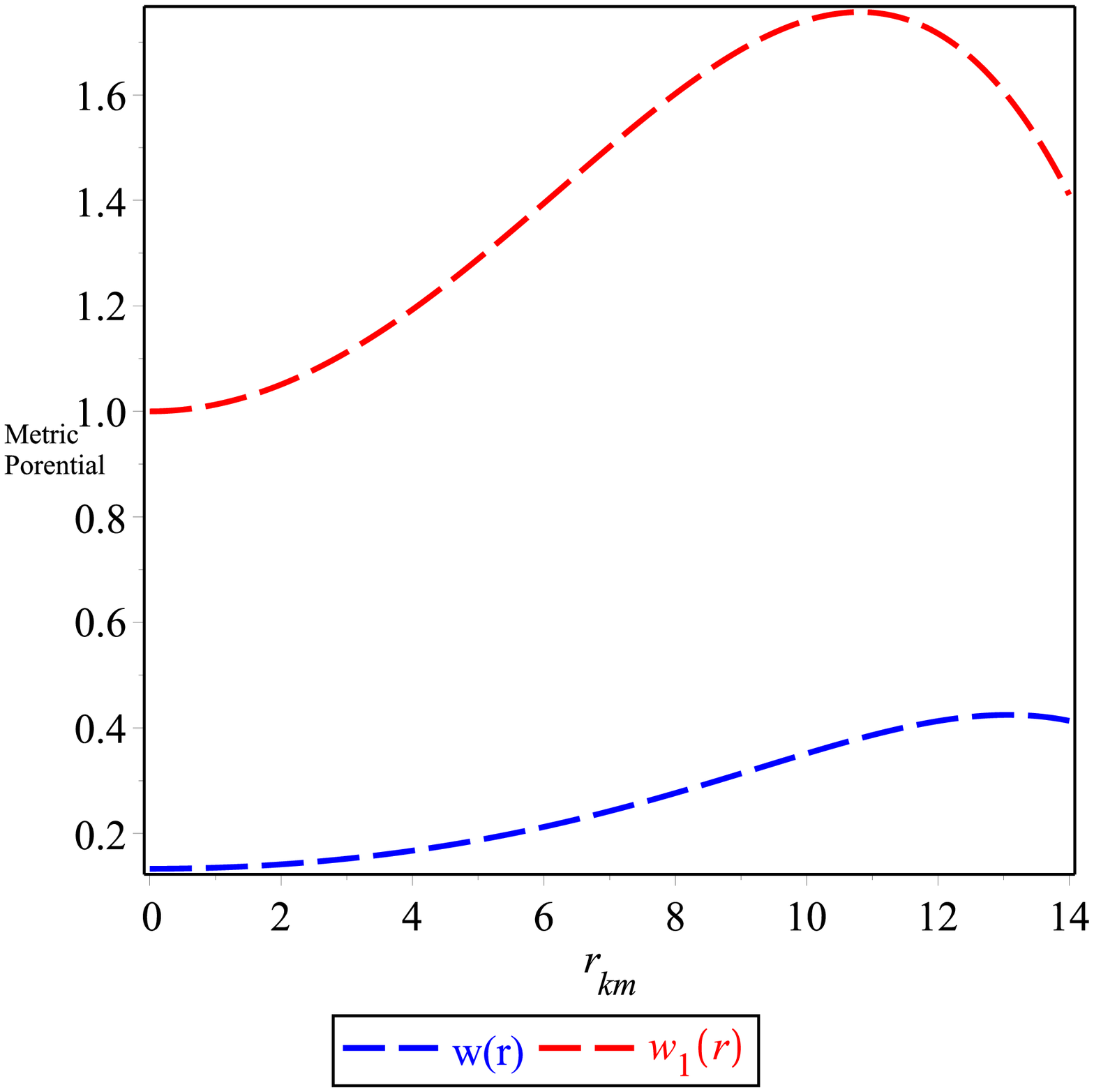}}
\subfigure[~Junction condition of the temporal component with Schwarzschild exterior solution]{\label{fig:junc}\includegraphics[scale=.3]{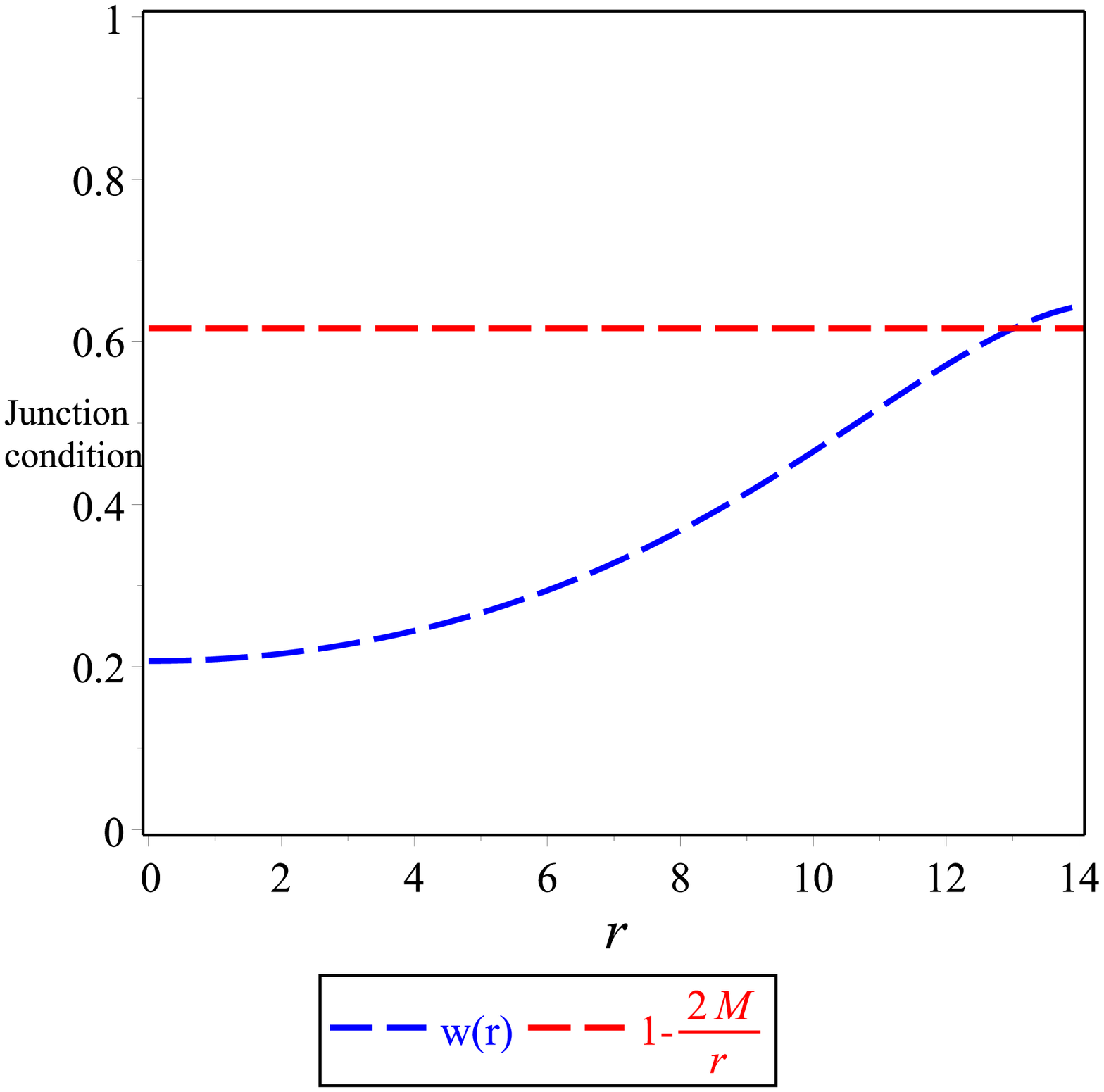}}
\subfigure[~Junction condition of the spatial component with Schwarzschild exterior solution]{\label{fig:junc1}\includegraphics[scale=.3]{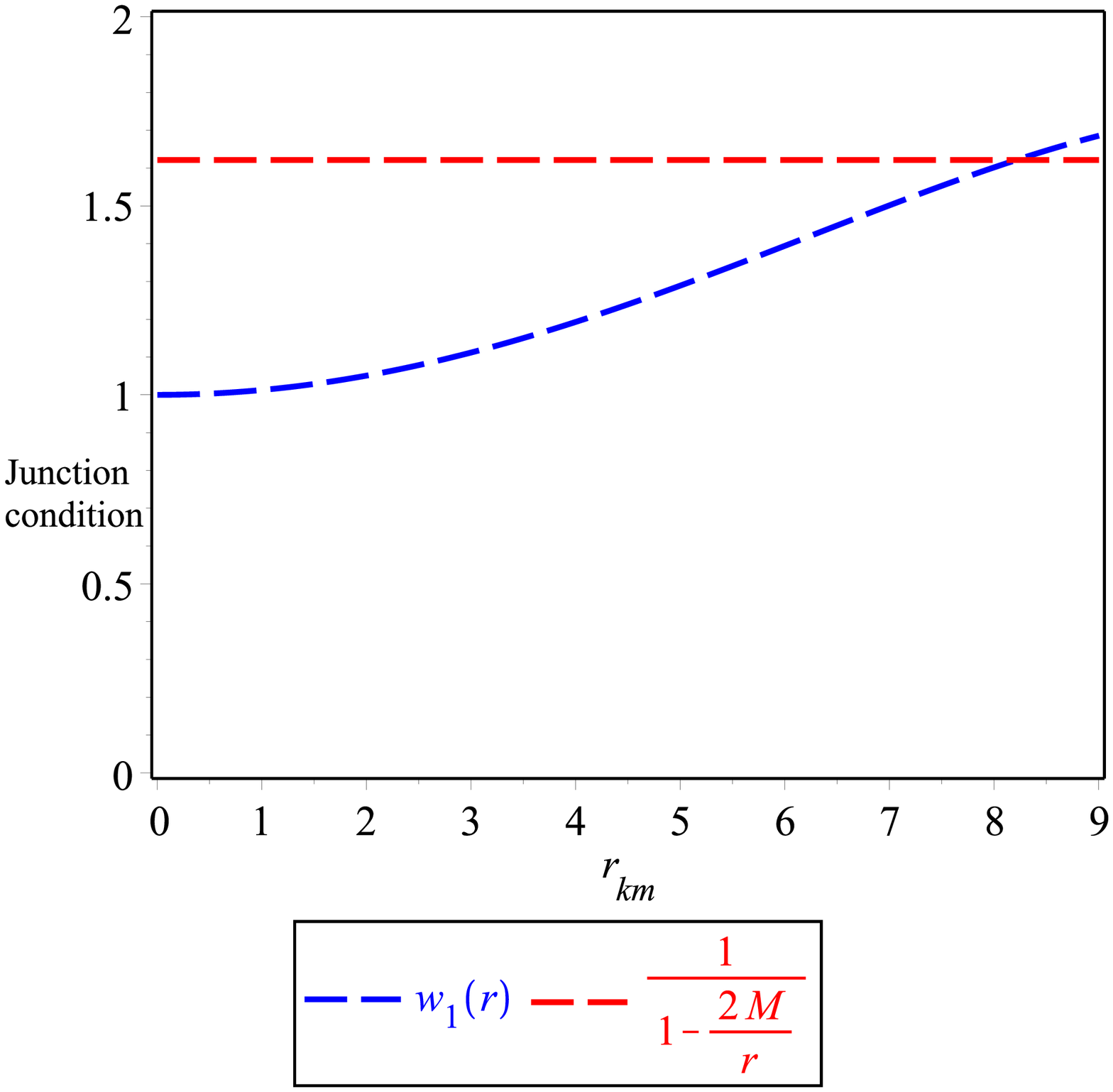}}
\caption[figtopcap]{\small{{Schematic plots: \subref{fig:pot1} the metric potentials (\ref{sol1}) and (\ref{sol11}); \subref{fig:junc} the junction condition at the surface boundary of star between the metric potential w(r) and the temporal  potential of Schwarzschild; \subref{fig:junc1} the junction condition at the surface boundary of star between the metric potential $w_1(r)$ and the spatial  potential of Schwarzschild.}}}
\label{Fig:1}
\end{figure}

Adopting these  constants, the physical quantities can be plotted.
The regular behavior of these can be assumed as being a first requirement to fit a realistic star model.  Figure \ref{Fig:1} \subref{fig:pot1}
represents  the  behavior of metric potentials for
\textrm{4U 1820-30} as well as the junction conditions. As Figure  \ref{Fig:1} shows,  the metric potentials assume the values  $w(0)=0.1328969066$ and $w_1(0)=1$ for $r=0$ and $s_3=0.5$. This means that they are both finite and positive at the center.

The metric potentials of  solution (\ref{psol}) are plotted in Figure \ref{Fig:1}  \subref{fig:pot1};  Figure \ref{Fig:1}  \subref{fig:junc} represents the junction condition of the temporal metric potential  (\ref{sol11}) and the temporal metric potential of the Schwarzschild solution.  Figure \ref{Fig:1} \subref{fig:junc1} represents the junction condition of the spatial  metric potential  (\ref{sol1}) and the spatial metric potential of the Schwarzschild  solution.
\begin{figure}
\centering
\subfigure[~Density]{\label{fig:density}\includegraphics[scale=0.3]{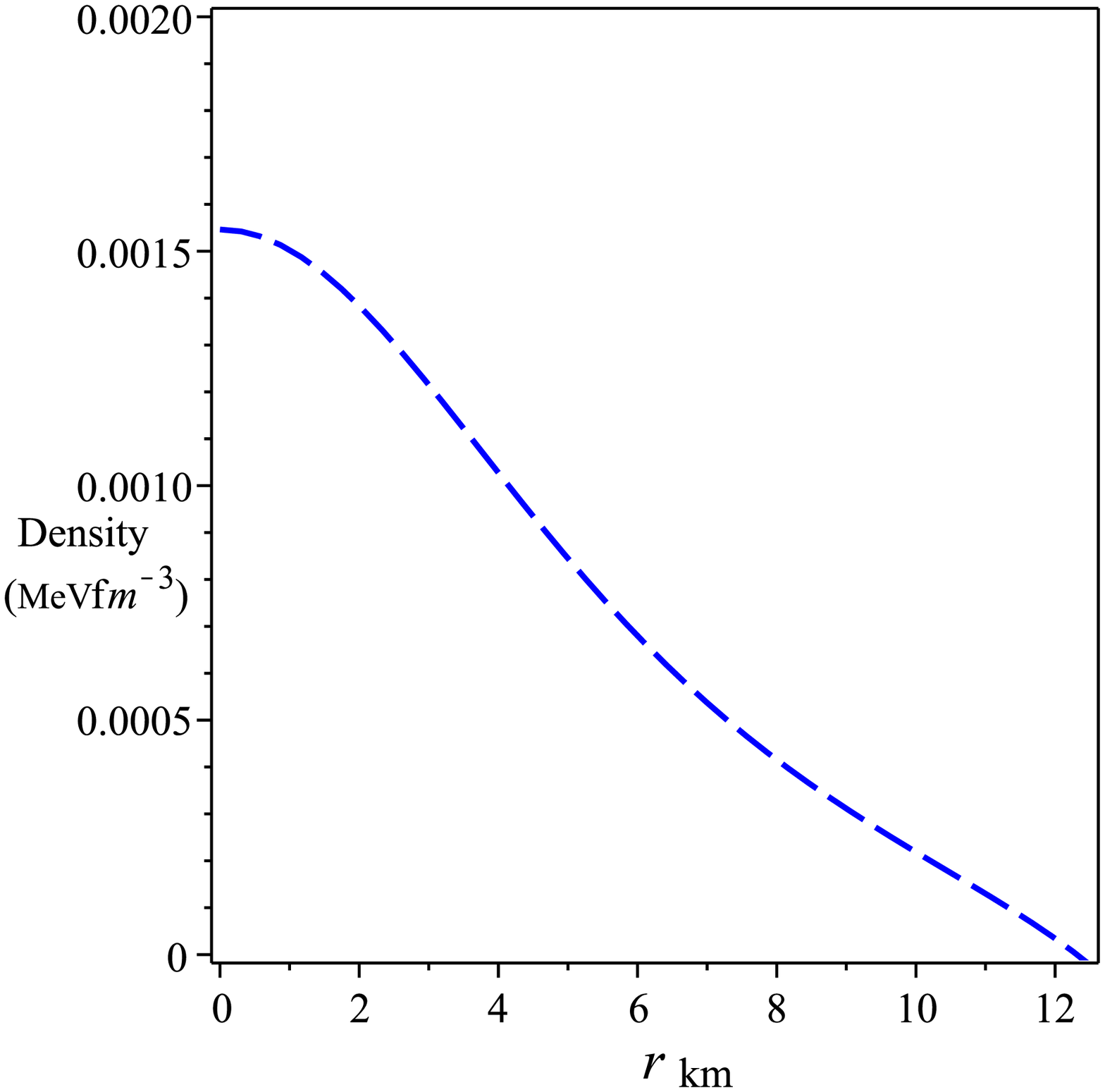}}
\subfigure[~Radial pressure]{\label{fig:pressure}\includegraphics[scale=.3]{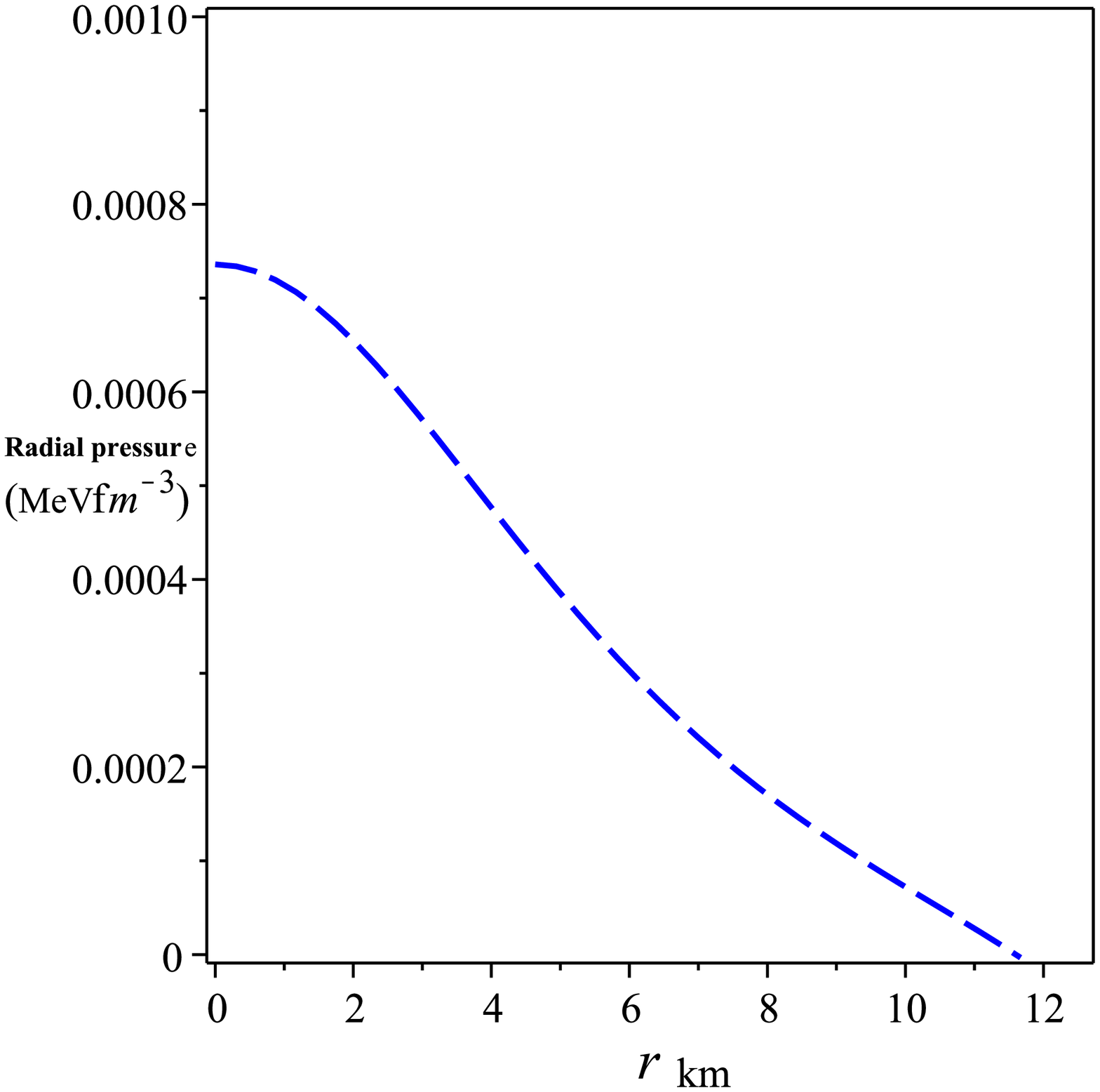}}
\subfigure[~Tangential pressure1]{\label{fig:pressuret}\includegraphics[scale=.3]{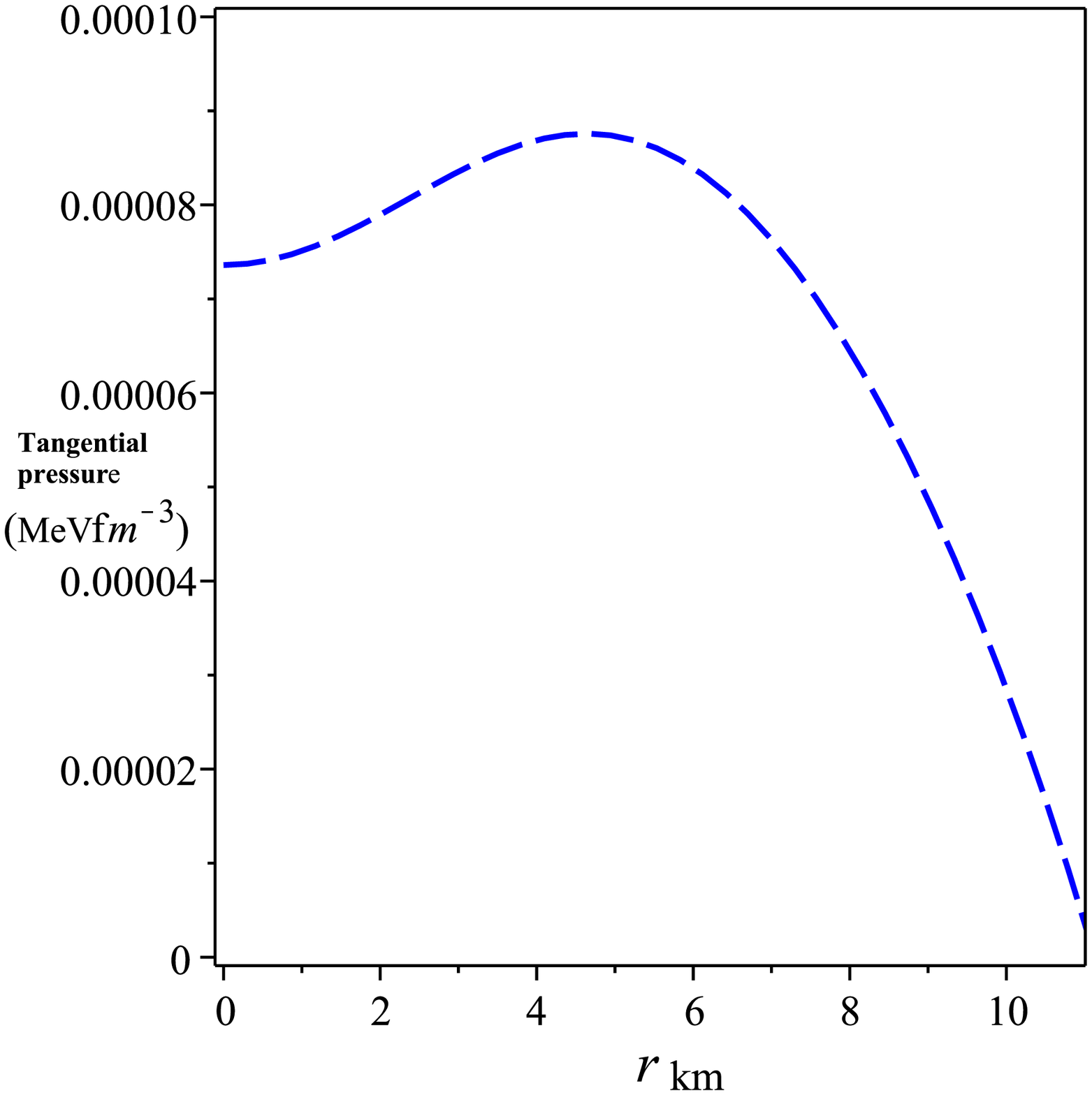}}
\caption[figtopcap]{\small{{Plot of the radial coordinate  $r$ in km  vs. the  density, radial and tangential pressures of (\ref{psol})  using the constants constrained from 4U 1820-30.}}}
\label{Fig:2}
\end{figure}
Figure  \ref{Fig:2} shows that density, radial and tangential pressures are positive as required for realistic  stellar configuration. Moreover, as Figure \ref{Fig:2} \subref{fig:density} shows, the density is high at the center, $\rho(r=0)=0.001546366234$ and  decreases  with the distance from it.  Figure \ref{Fig:2} \subref{fig:pressure}  shows that the radial pressure  tends to be zero at the boundary; which  is relevant for a realistic model. Figure \ref{Fig:2} \subref{fig:pressuret} shows that the tangential pressure has a positive value and that it has a high value at the center, decrease away from the center.\\
\begin{figure}
\centering
\subfigure[~Anisotropy and Anisotropic force]{\label{fig:An}\includegraphics[scale=0.3]{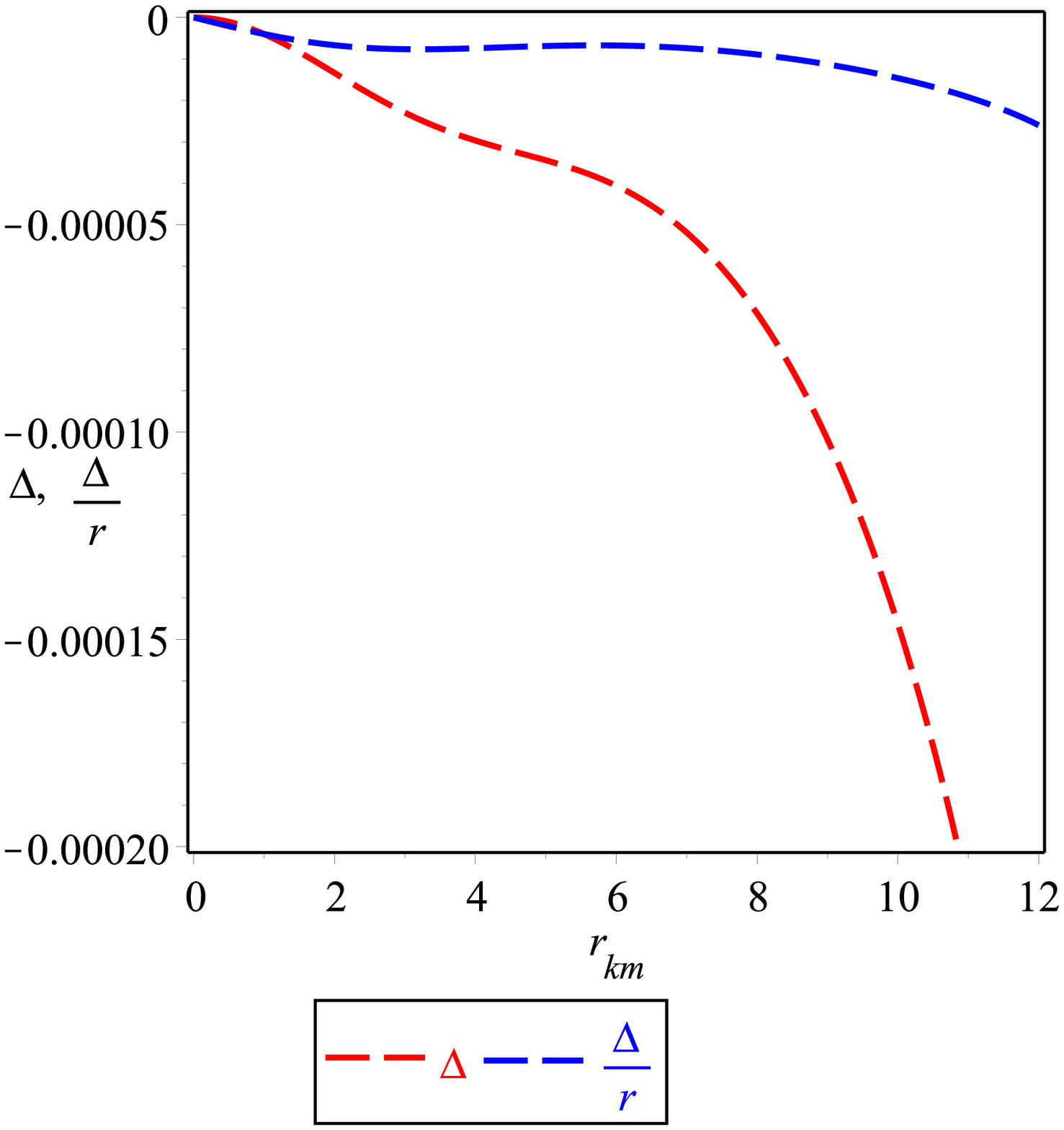}}
\subfigure[~The gradient of energy-density, radial and tangential pressure]{\label{fig:grad}\includegraphics[scale=.3]{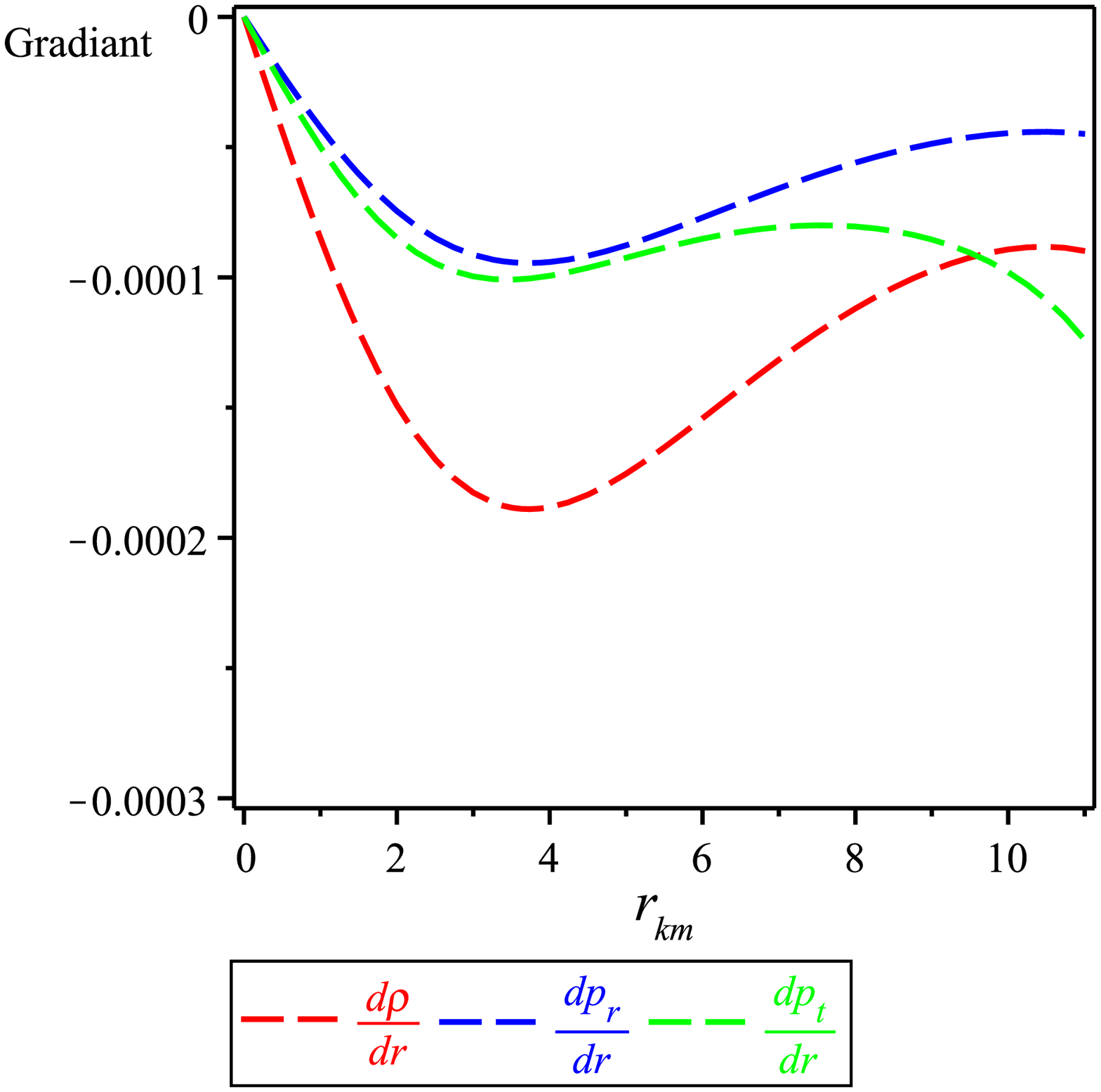}}
\subfigure[~Equation of state]{\label{fig:EoS}\includegraphics[scale=.3]{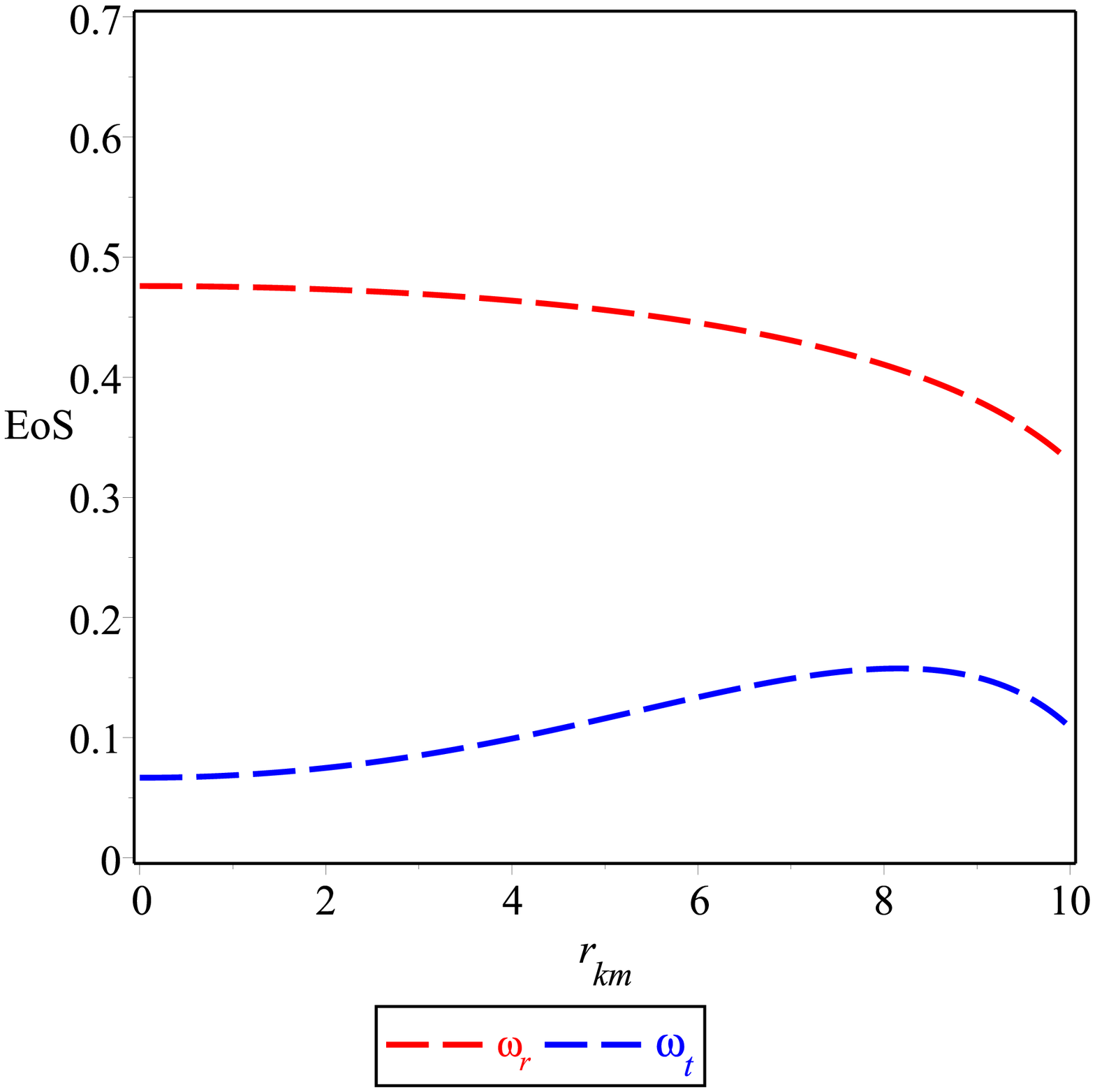}}
\caption[figtopcap]{\small{{Schematic plots: \subref{fig:An} of the radial coordinate  $r$ in km  vs. the  anisotropy and anisotropic force of solution (\ref{psol})  using the constants constrained from  4U 1820-30; \subref{fig:grad} the radial coordinate  $r$ in km  vs. the  gradient of density, radial and tangential pressures of solution (\ref{psol})  using the constants constrained from 4U 1820-30; \subref{fig:EoS} the radial and tangential equation of states using the constants constrained from 4U 1820-30.}}}
\label{Fig:3}
\end{figure}
Figure  \ref{Fig:3} \subref{fig:An} shows that the anisotropy, $\Delta(r)=p_t-p_r$ and the anisotropic force. As Figure \ref{Fig:3} \subref{fig:An} shows, the anisotropy vanishs at the center and decreases at the surface of the star. In particular,  Fig.  \ref{Fig:3} \subref{fig:An} shows that the anisotropic force $\frac{\Delta}{r}$ is negative.  This means that it  possesses an inward gravitational since  $p_r-p_t>0$. Figure \ref{Fig:3}  \subref{fig:grad} shows that the gradients of  density, radial and tangential pressures are negative  confirming the decreasing of density, radial and transverse pressures through the stellar configuration. Also Fig. \ref{Fig:3} \subref{fig:EoS} shows that the radial and tangential EoSs are positive within the star configuration.
\begin{figure}
\centering
\subfigure[~Weak  energy conditions]{\label{fig:WEC}\includegraphics[scale=0.3]{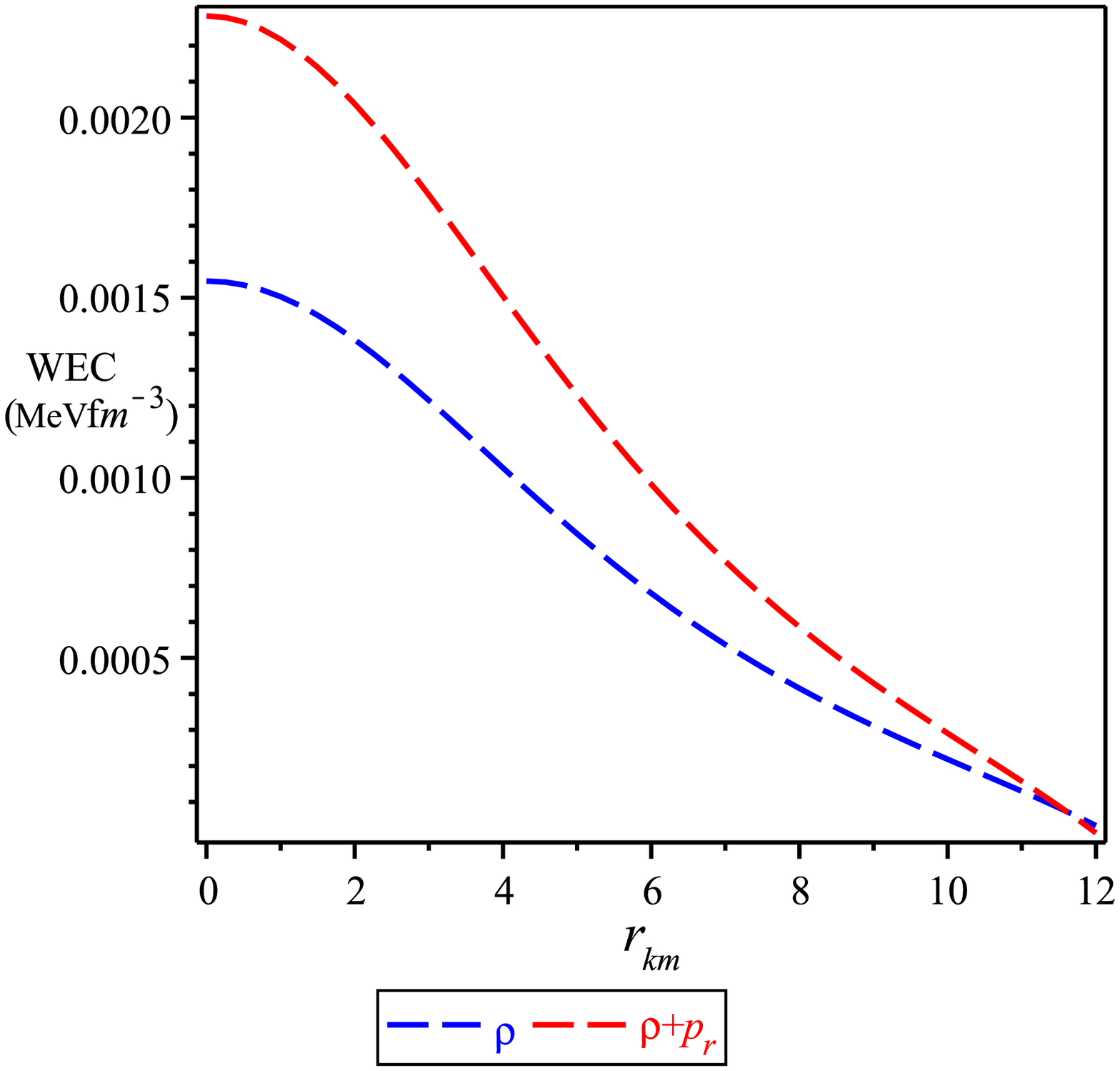}}
\subfigure[~Null  energy conditions]{\label{fig:NEC}\includegraphics[scale=0.3]{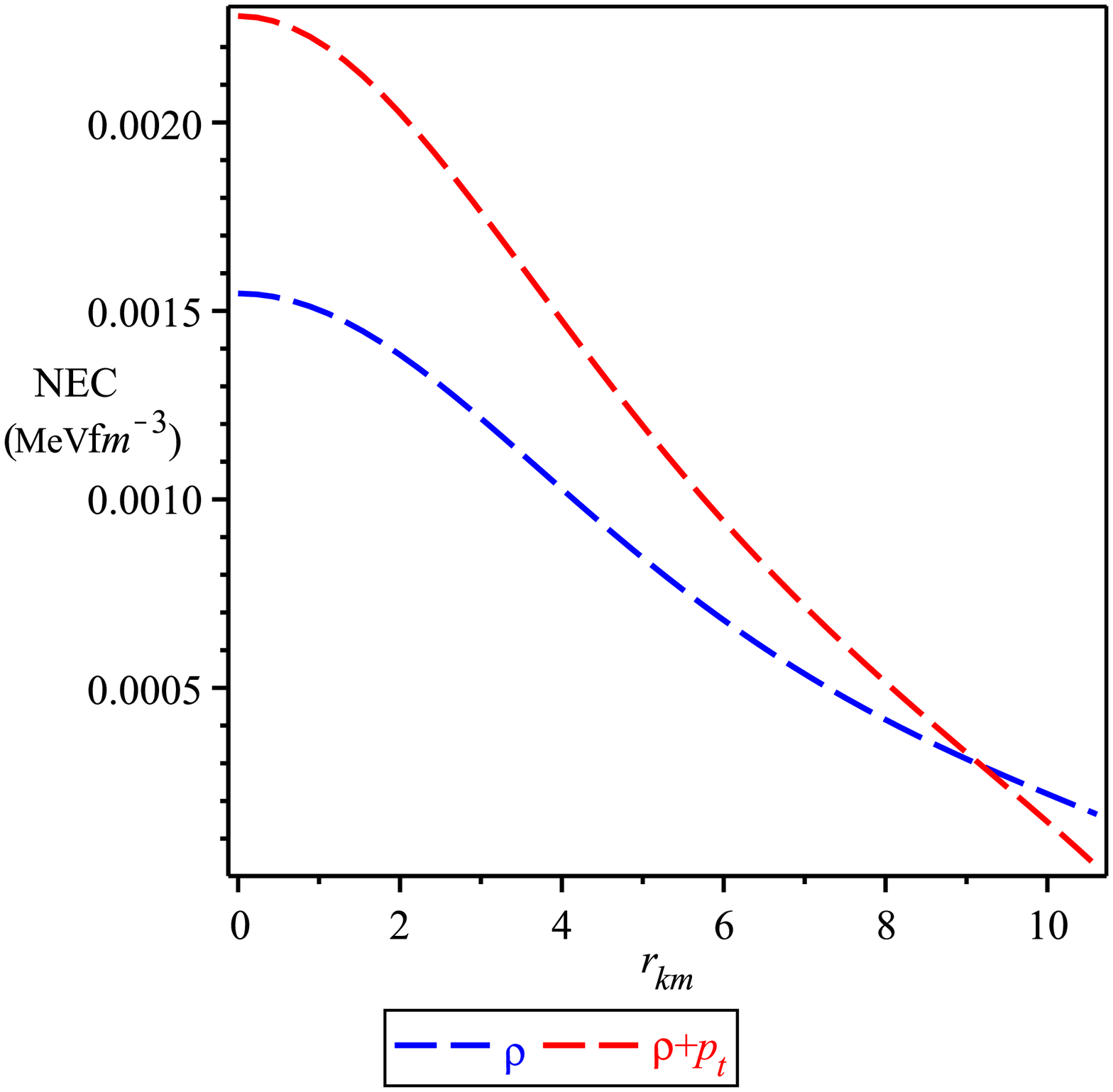}}
\subfigure[~Strong energy condition]{\label{fig:SEC}\includegraphics[scale=.3]{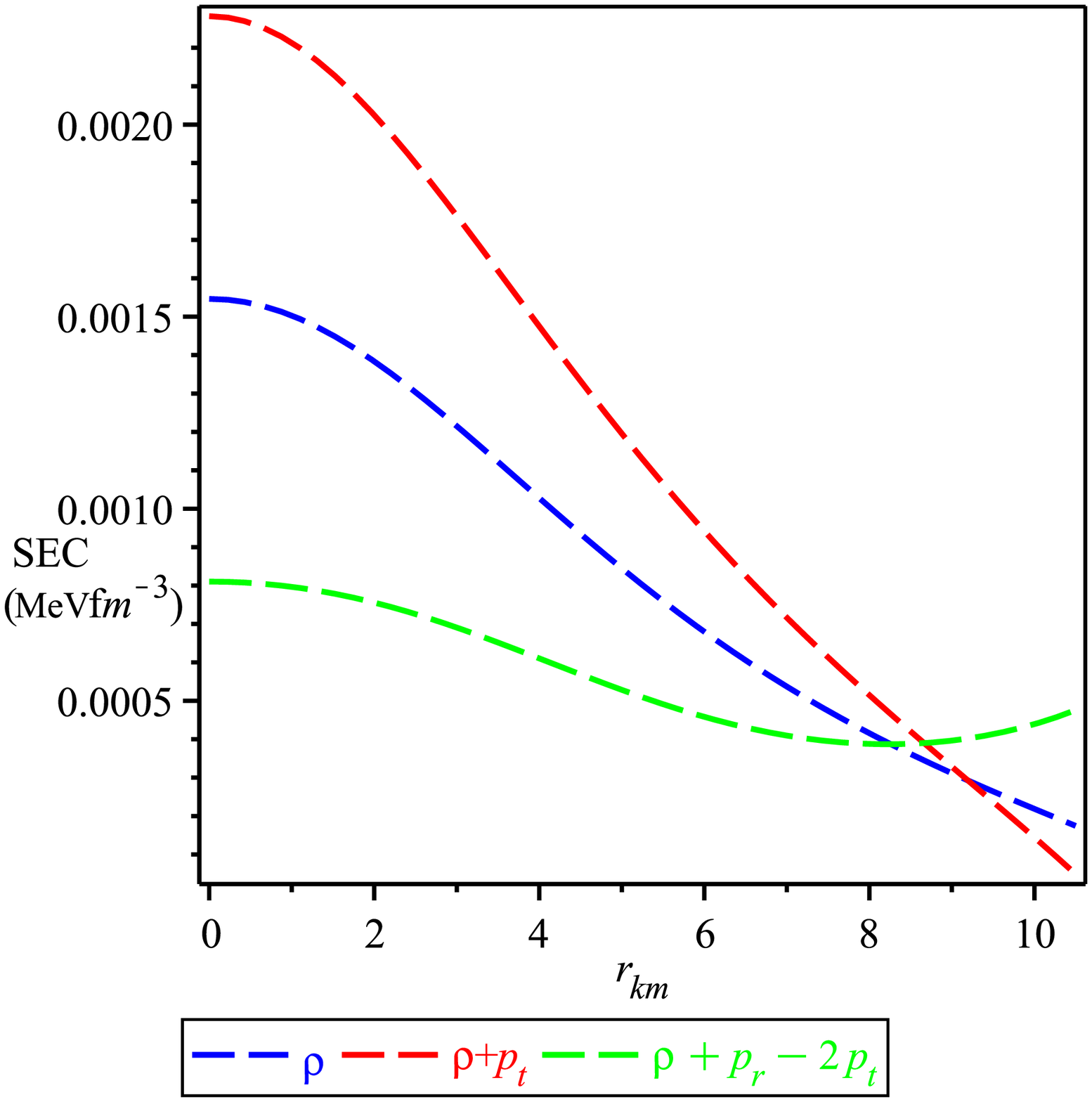}}
\subfigure[~Dominant energy condition]{\label{fig:DEC}\includegraphics[scale=.3]{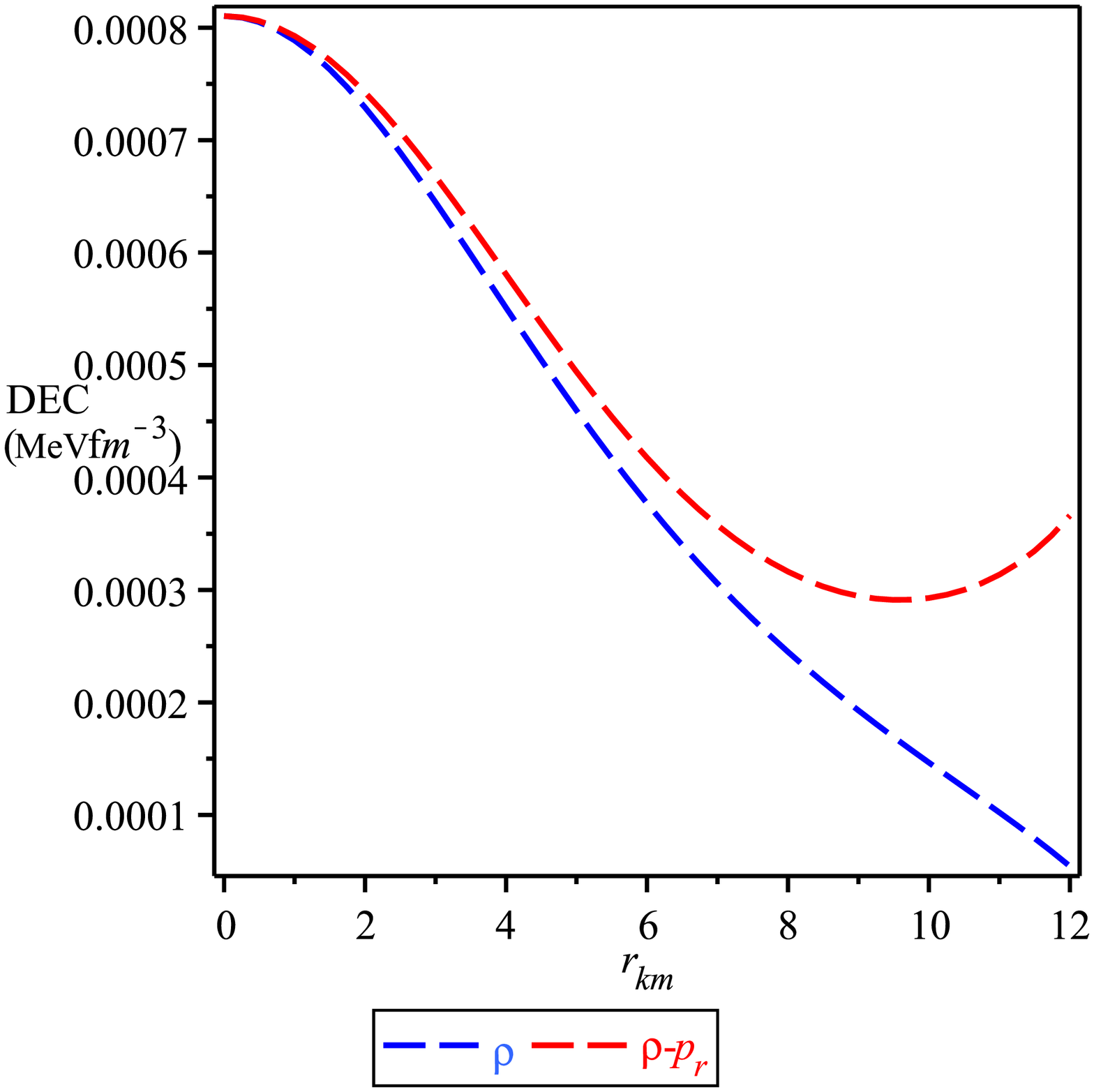}}
\caption[figtopcap]{\small{{Schematic plots:  \subref{fig:WEC} the  weak, \subref{fig:NEC} null, \subref{fig:SEC} strong and \subref{fig:DEC} dominant energy conditions of solution (\ref{psol})  using the constants constrained from 4U 1820-30.}}}
\label{Fig:4}
\end{figure}

  Figures \ref{Fig:4} \subref{fig:WEC}, \ref{Fig:4} \subref{fig:NEC},   \ref{Fig:4} \subref{fig:SEC} and \ref{Fig:4} \subref{fig:DEC} show the positive values of the WEC,  NEC,  SEC and DEC.   Therefore, all the energy conditions are satisfied throughout the stellar configuration as required for a physically meaningful stellar model.

\begin{figure}
\centering
\subfigure[~Radial and tangential speed of sound ]{\label{fig:speed}\includegraphics[scale=0.3]{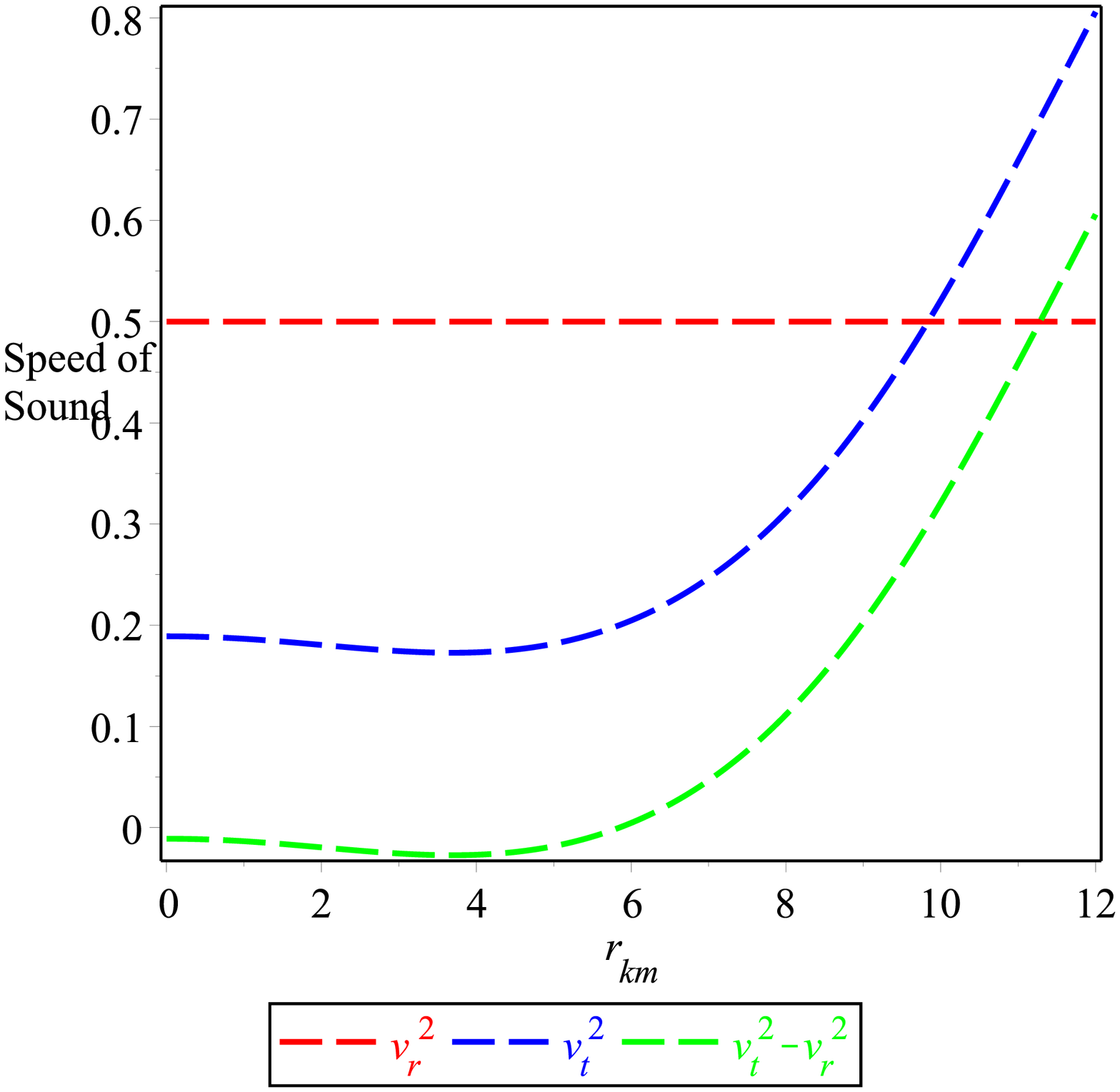}}
\subfigure[~Mass function]{\label{fig:mass}\includegraphics[scale=0.3]{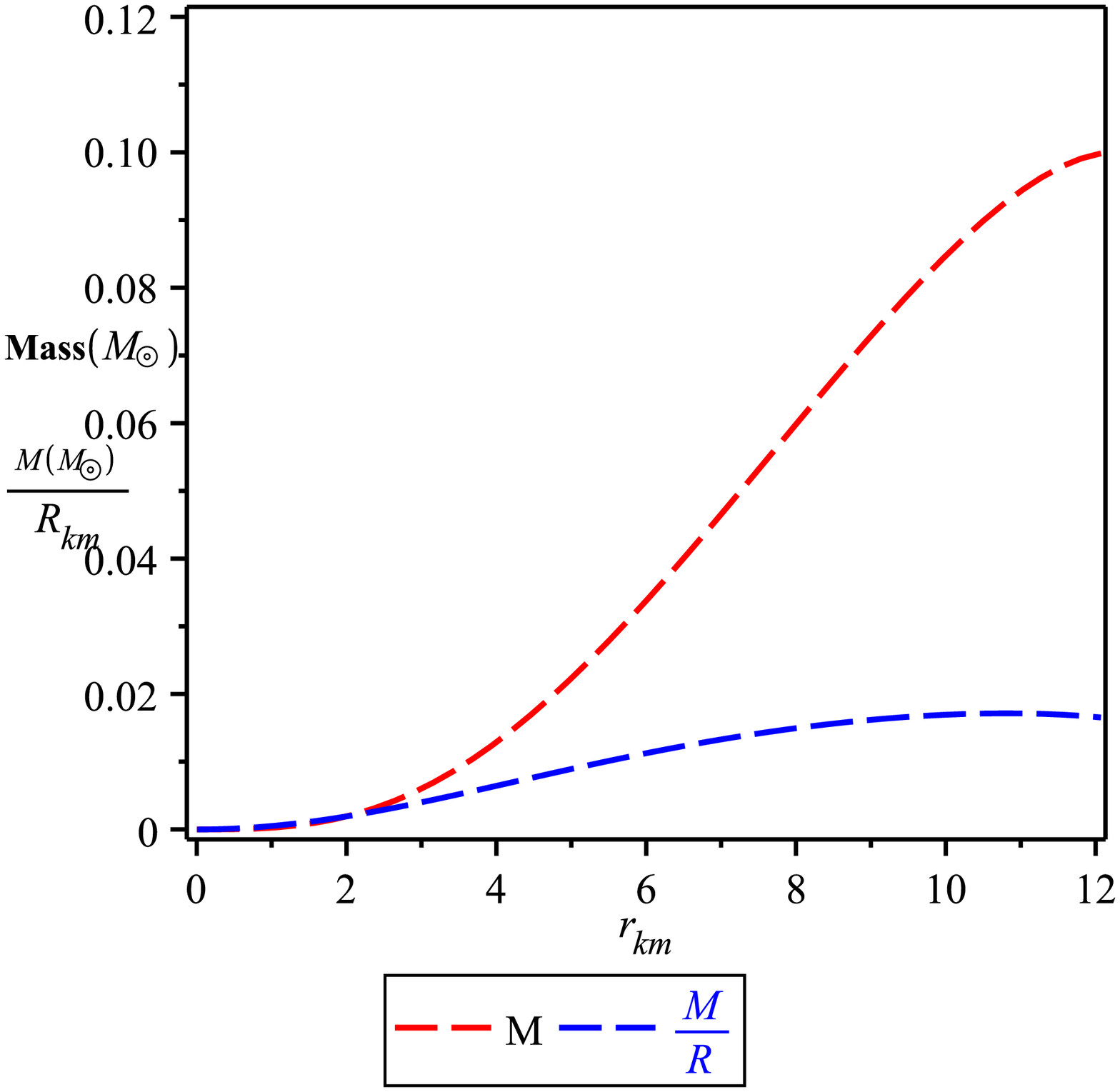}}
\subfigure[~Redshift ]{\label{fig:redshift}\includegraphics[scale=0.3]{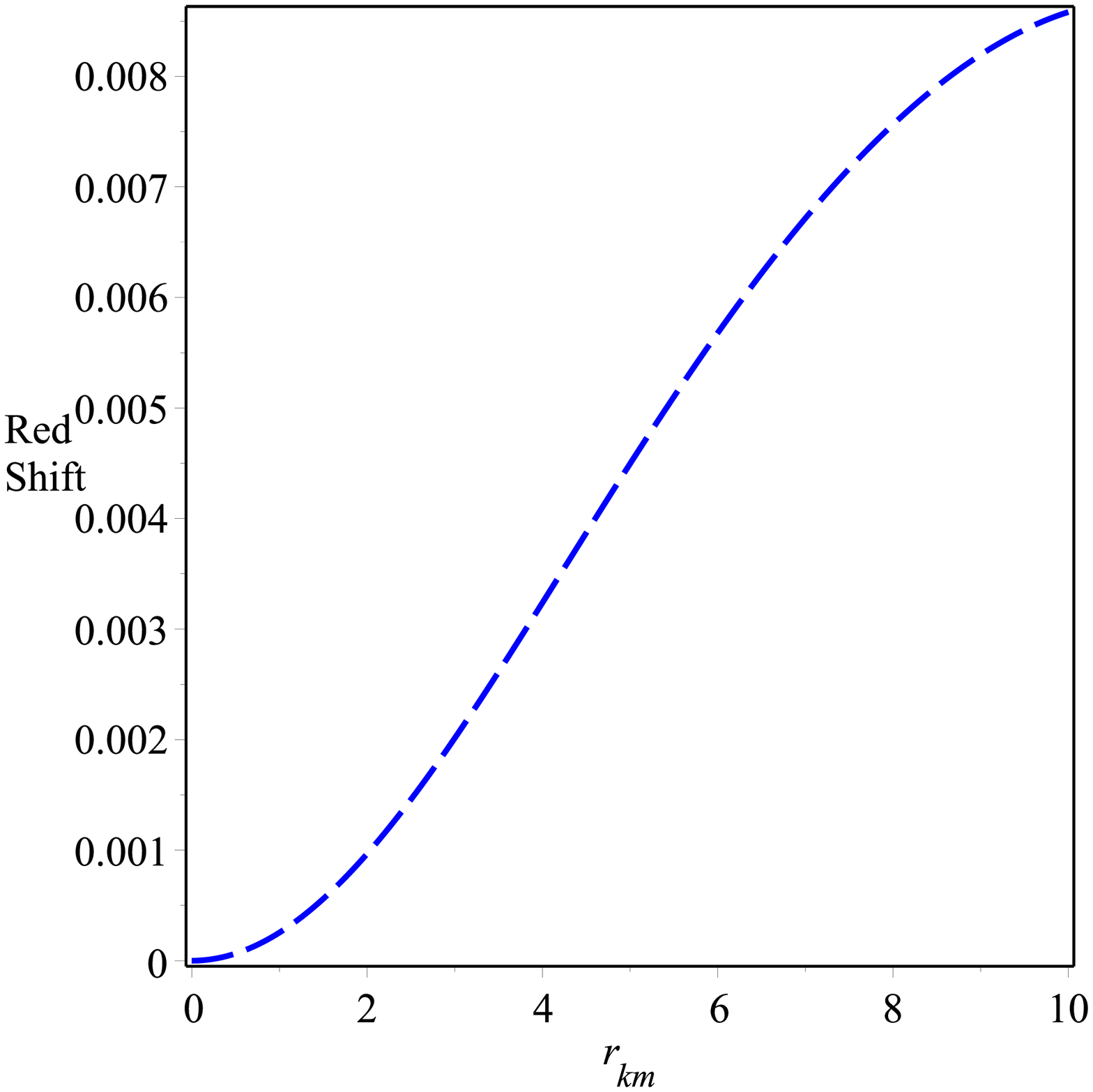}}
\caption[figtopcap]{\small{{Schematic plots: \subref{fig:speed} of the radial coordinate  $r$ in km  vs, the radial,  tangential speeds of solution (\ref{psol}) and the difference between  the radial and tangential speeds of sound; \subref{fig:mass} the mass and compactness vs the radial coordinate  $r$ in km  \subref{fig:redshift} the surface red-shift vs the radius $r$ using the constants constrained from 4U 1820-30.}}}
\label{Fig:5}
\end{figure}

Figure \ref{Fig:5}  \subref{fig:speed} represents  the velocity of
the sound, radial and transverse, speeds; this should be less than the speed
of the light, i.e., $\frac{dpr}{d\rho}<1$ and $\frac{dpt}{d\rho}<1$. This condition is known as the causality
condition.
The mass function  given  by Eq. (\ref{mas1}) is plotted in Fig. \ref{Fig:5}\subref{fig:mass}, showing that it is a monotonically
increasing function of the radial coordinate and $M(r=0) = 0$.  Furthermore, Figure \ref{Fig:5}\subref{fig:mass} shows  the behavior of the compactness parameter of star which is increasing. The radial variation of the surface redshift is plotted in  Figure \ref{Fig:5}\subref{fig:redshift}.
 B\"{o}hmer and Harko \citep{Bohmer2006} constrained the surface red-shift to be  $Z\leq 5$. The surface redshift of this model is calculated according to $4U 1820-30$ and is found to be $\approx 0.008$.

\section{Stability of the model}\label{stability}
In this section we are going to discuss the stability issue using two different techniques; the TOV equations and the adiabatic index.
\subsection{Equilibrium analysis through TOV equation}
\begin{figure}
\centering
\subfigure[~Different forces]{\label{fig:TOV}\includegraphics[scale=0.3]{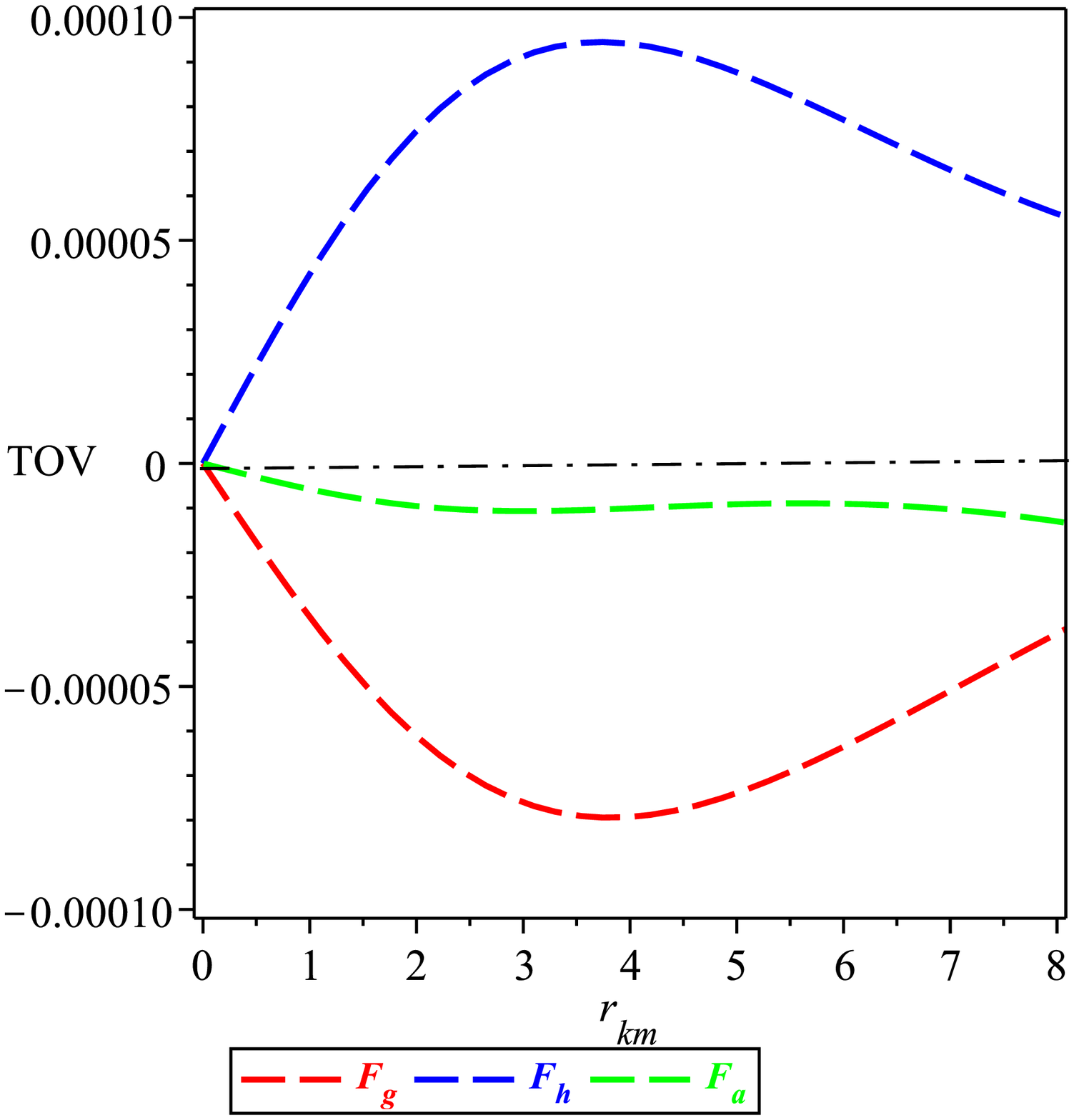}}
\subfigure[~Adiabatic index]{\label{fig:adi}\includegraphics[scale=0.3]{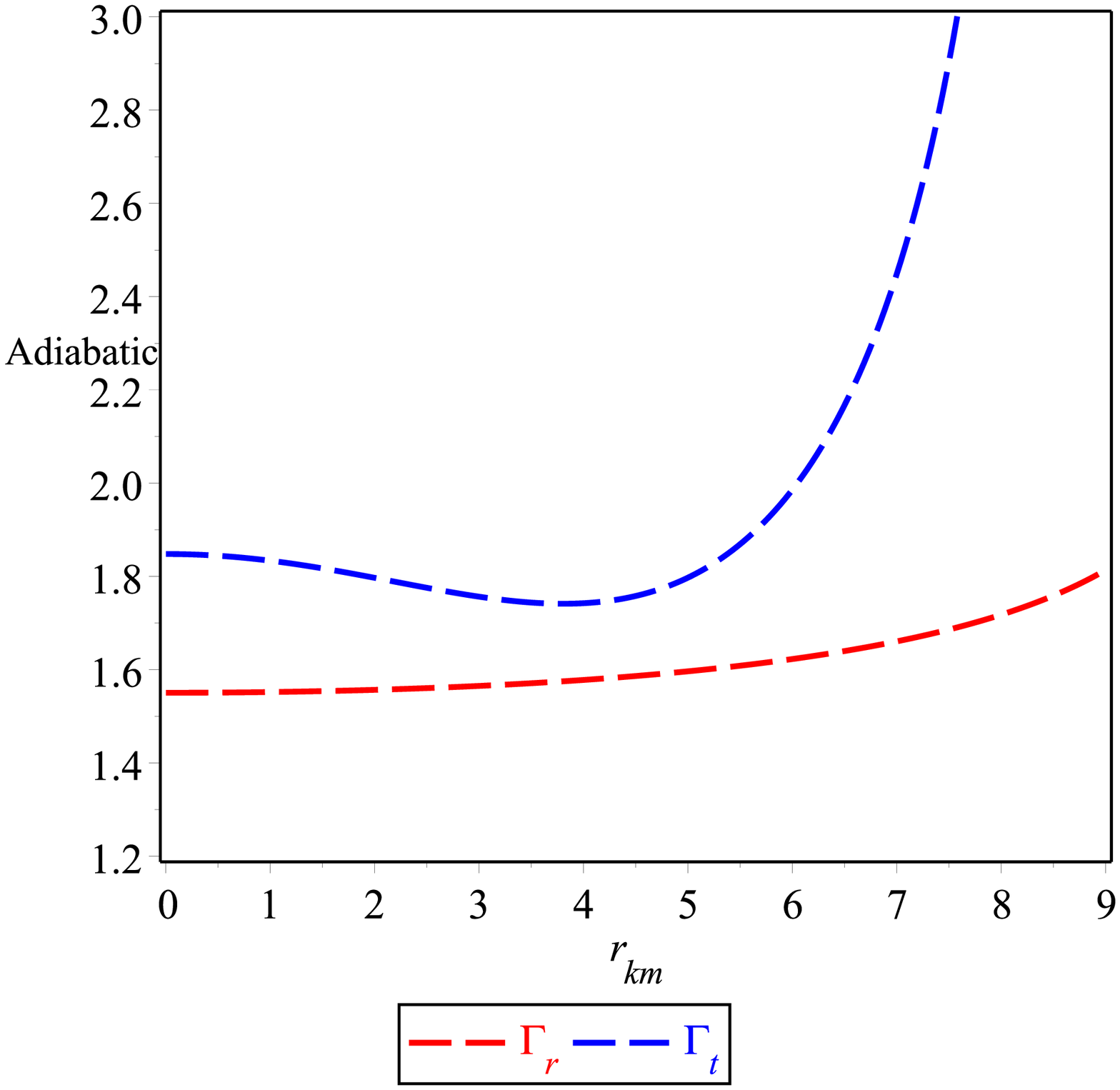}}
\caption[figtopcap]{\small{{Schematic plots: \subref{fig:TOV} the gravitational,  the anisotropic and  the hydrostatic  forces  vs. the  radius  $r$; \subref{fig:adi} the adiabatic index vs. the radius $r$ using the constants
constrained from  4U 1820-30}.}}
\label{Fig:6}
\end{figure}

In this subsection, we are going to discuss the stability of
any stellar  model. For this goal,  we assume hydrostatic equilibrium
through the TOV equation.
Using the TOV equation \citep{PhysRev.55.364,PhysRev.55.374} as that presented  in \citep{PoncedeLeon1993}, we obtain the following  form:
\begin{eqnarray}\label{TOV}   \frac{2[p_t-p_r]}{r}-\frac{M_g(r)[\rho(r)+p_r]\sqrt{w}}{r\sqrt{w_1}}-\frac{dp_r}{r}=0,
 \end{eqnarray}
with $M_g(r)$ being  the gravitational mass at
radius $r$, as defined by the  Tolman-Whittaker mass
formula which gives:
\begin{eqnarray}\label{ma}   M_g(r)=4\pi{\int_0}^r\Big({T_t}^t-{T_r}^r-{T_\theta}^\theta-{T_\phi}^\phi\Big)r^2\sqrt{ww_1}dr=\frac{rw'\sqrt{w_1}}{2w}\,,
 \end{eqnarray}
Inserting Eq. (\ref{ma}) into (\ref{TOV}),  we obtain:
\begin{eqnarray}\label{ma1}  \frac{2(p_t-p_r)}{r}-\frac{dp_r}{dr}-\frac{w'[\rho+p_r]}{2\sqrt{w}}=F_g+F_a+F_h=0\,,
 \end{eqnarray}
 where $F_g=-\frac{w'[\rho+p_r]}{2\sqrt{w}}$,  $F_a=\frac{2(p_t-p_r)}{r}$, and $F_h=-\frac{dp_r}{dr}$  are the gravitational,  anisotropic and hydrostatic forces,  respectively. The behavior of the TOV equation for model  (\ref{psol}) is shown in Figure \ref{Fig:6} in which the three different forces are plotted.
 This shows that hydrostatic force is positive and is dominated by the gravitational and anisotropic forces, which are negative to maintain the system in static equilibrium.

\subsection{Adiabatic index}
The stable equilibrium configuration of a spherically symmetric system can be studied using the adiabatic index, which is a basic ingredient of the stability criterion. Let us consider an adiabatic perturbation, the adiabatic index $\Gamma$, is defined as \citep{1964ApJ...140..417C,Nashed:2001im,1989A&A...221....4M,10.1093/mnras/265.3.533}:
\begin{eqnarray}\label{a11}  \Gamma=\left(\frac{\rho+p}{p}\right)\left(\frac{dp}{d\rho}\right)\,.
 \end{eqnarray}
 A Newtonian isotropic sphere is in stable
equilibrium if the adiabatic index $\Gamma>\frac{4}{3}$
as reported in  Heintzmann
and Hillebrandth \citep{1975A&A....38...51H}. For $\Gamma=\frac{4}{3}$, the  isotropic
sphere is  in neutral equilibrium. Based on some  works carried out by Chan et al. \citep{10.1093/mnras/265.3.533},  the following
condition is required for the stability of a relativistic anisotropic sphere $\Gamma >\gamma$ where: \begin{eqnarray}\label{ai}  \gamma=\frac{4}{3}-\left\{\frac{4(p_r-p_t)}{3| p'_r|}\right\}_{max}\,.
 \end{eqnarray}
 Using Eq. (\ref{ai}),  we obtain:
 \begin{eqnarray}\label{a12}  &&\Gamma=\frac{4}{3}-\frac{2}{3}\Big[(c_2r^2+1)^2(4k_1{}^2k^8+6k^4-8r^2k^2+3r^4)\Big]\times\Big[2(c_2r^2+1)^2k^8k_1{}^2+16k^6c_2-18c_2{}^2r^4k^4-36k^4c_2r^2+6k^4\nonumber\\
 &&+16c_2r^4k^2-16k^2r^2+16c_2{}^2r^6k^2+2c_2r^6+9r^4-3c_2{}^2r^8+4c_2{}^2k^8\Big]^{-1}.
 \end{eqnarray}
 From Eq. (\ref{a11}),  we obtain  the adiabatic index of solution (\ref{psol})   in the form:
 \begin{eqnarray}\label{aic} &&\Gamma_r=s_3\, \Bigg( 8\,s_4\,\pi\,{r}^{8}{s_2}^{2}+16\,{r}^{6}s_4
\,\pi\,s_1s_2+{r}^{6}s_3\,{s_2}^{2}+8\,{r}^{4}s_4\,\pi\,{s_1}^{2}+2\,{r}^
{4}s_2s_3\,s_1+16\,s_4\,\pi\,{r}^{4}s_2+{r}^{2}s_3\,{s_1}^{2}+16\,
s_4\,\pi\,{r}^{2}s_1\nonumber\\
&&+5\,s_3\,{r}^{2}s_2+3\,s_3\,s_1+8\,s_4\,\pi+{s_2
}^{2}{r}^{6}+2\,s_1{r}^{4}s_2+5\,s_2{r}^{2}+{r}^{2}{s_1}^{2}+3\,s_1 \Bigg)\times\Bigg\{8
\,s_4\,\pi\,{r}^{8}{s_2}^{2}+16\,{r}^{6}s_4\,\pi\,s_1s_2+{r}^{6}s_3\,
{s_2}^{2}\nonumber\\
&&
+8\,{r}^{4}s_4\,\pi\,{s_1}^{2}+2\,{r}^{4}s_2s_3\,s_1+16\,s_4\,
\pi\,{r}^{4}s_2+{r}^{2}s_3\,{s_1}^{2}+16\,s_4\,\pi\,{r}^{2}s_1+5\,
s_3\,{r}^{2}s_2+3\,s_3\,s_1+8\,s_4\,\pi\Bigg\}^{-1}
\,.
 \end{eqnarray}
In Fig. \ref{Fig:6}\subref{fig:adi}  $\Gamma_r$ and  $\Gamma_t$  are  reported.
As can be seen from these plots, the value of  $\Gamma_t$  is greater
than that of $\Gamma_r$ throughout the stellar interior, and hence, the stability condition is fulfilled.

\subsection{Stability in the static state}
\begin{figure}
\centering
\subfigure[~Mass as a function of central density]{\label{fig:mrho}\includegraphics[scale=0.3]{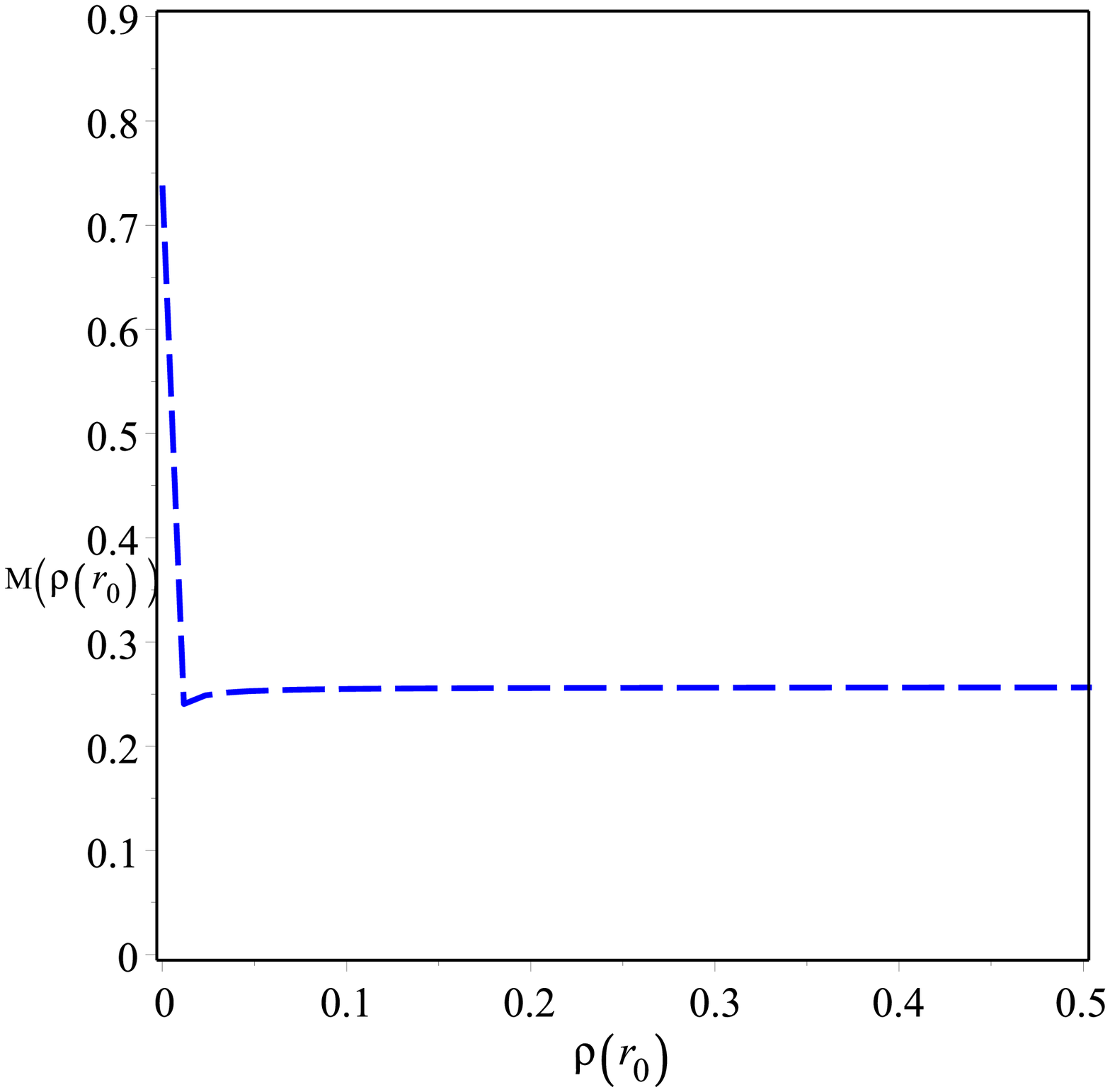}}
\subfigure[~differentiation of mass w.r.t. the central density]{\label{fig:dmrho}\includegraphics[scale=0.3]{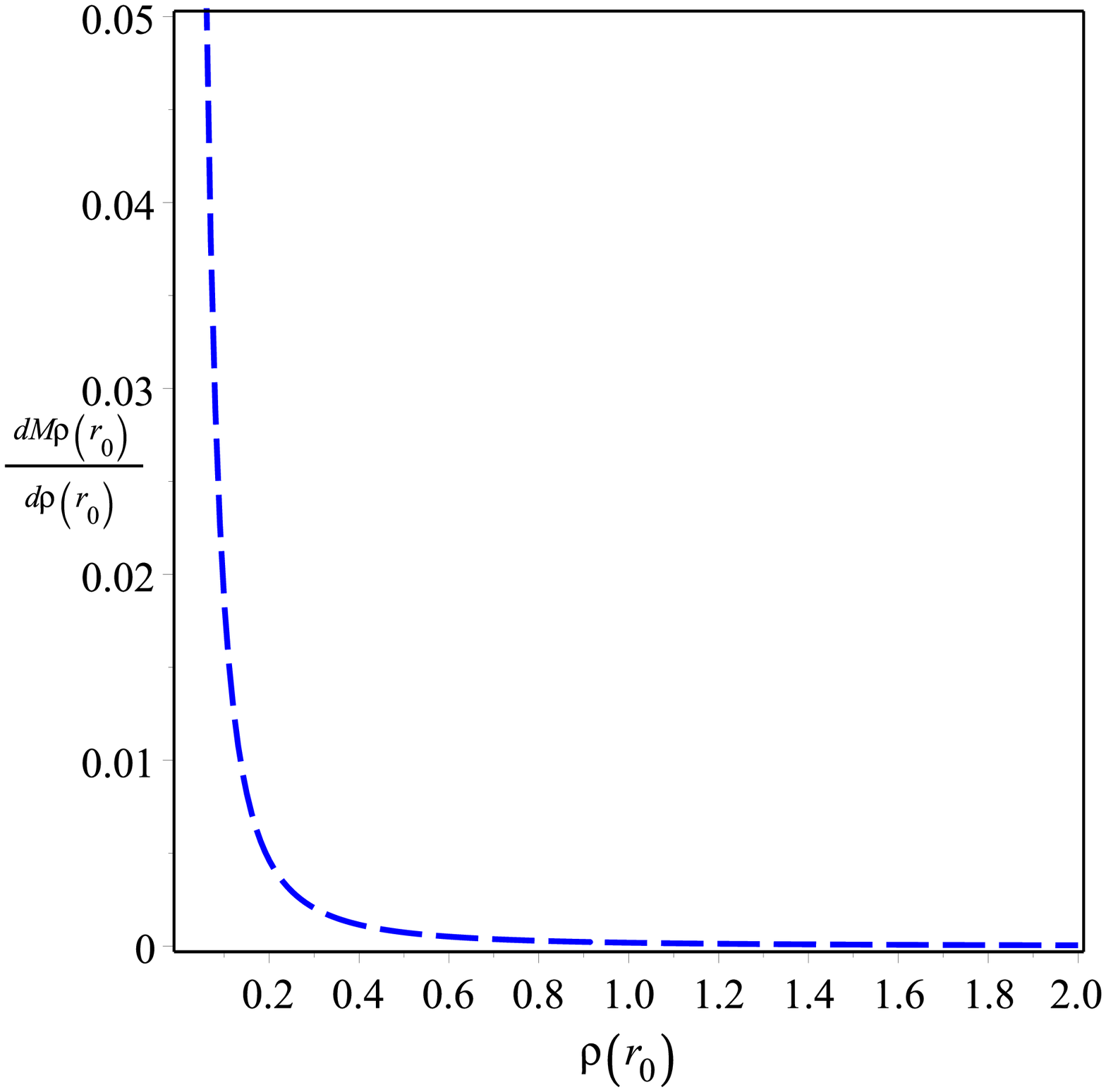}}
\caption[figtopcap]{\small{{Schematic plots: \subref{fig:mrho} the gravitational mass as a function of the central density; \subref{fig:dmrho} the differentiation of gravitational mass w.r.t. the central density using the
constrained from  4U 1820-30}.}}
\label{Fig:7}
\end{figure}

For stable compact stars in terms of the mass-central mass--radius, and relations for the energy density, Harrison, Zeldovich, and Novikov  \citep{1965gtgc.book.....H,1971reas.book.....Z} claimed that the gradient of the central density, with respect to the mass, must be positive, i.e., $\frac{\partial M}{\partial \rho_{r_0}}> 0$. If this condition is satisfied, then we have stable configurations.  To be more specific, the stable stable or unstable region is satisfied
for constant mass i.e. $\frac{\partial M}{\partial \rho_{r_0}}= 0$ \citep{Singh:2019ykp}. Let us apply this procedure to our solution (\ref{psol}).   For  solution (\ref{psol}), the  central density has the form:
\begin{eqnarray} \label{sta} &&\rho_{_{r_0}}=\frac{3s_1}{8\pi } \Rightarrow s_1=\frac{8\pi\rho_{r_0}}{3},\nonumber\\
 && M(\rho_{r_0})=\frac{R^3(8\,\pi\,\rho_{r_0}+3s_2 R^2)}{16\,\pi\,(3+8\pi R^2\rho_{r_0}+3 s_2 R^4)}\,.
 \end{eqnarray}
With Eq. (\ref{sta}) we have:
 \begin{eqnarray} \label{sta1} \frac{\partial M}{\partial \rho_{r_0}}=\frac{3 R^3}{2(3+8\pi R^2\rho_{r_0}+3 s_2 R^4)^2} \,.
 \end{eqnarray}
From Eq. (\ref{sta1})), the solution  (\ref{psol}) has a stable configuration since   $\frac{\partial M}{\partial \rho_{r_0}}> 0$ \citep{Singh:2019ykp}. The behaviors of (\ref{sta}) and (\ref{sta1}) are shown in Fig. \ref{Fig:7}. It follows from these figures that the mass and the gradient-of-mass decrease as the energy density become larger. The above discussion of solution  (\ref{psol}) shows that we have a good model.
\begin{table*}[t!]
\caption{\label{Table1}%
Values of model parameters \citep{2016ApJ...820...28O}}
\begin{ruledtabular}
\begin{tabular*}{\textwidth}{lccccccc}
{{Pulsar}}                              & Mass ($M_{\odot}$) &      {Radius (km)} &   {$s_1$}  &    {$s_2$}    & {$s_3$} & {$s_4$}&{$s_5$}\\ \hline
&&&&&&&\\
4U 1724-207       &$1.81^{+0.25}_{-0.37}$    & $12.2^{+1.4}_{-1.4}$ &   $\approx$0.013    &   $\approx$-0.45$\times10^{-4}$         &0.5 &$\approx$-0.18$\times10^{-12}$&$\approx$0.10    \\
4U 1820-30 & $1.46^{+0.21}_{-0.21}$ & $11.1^{+1.8}_{-1.8}$ &   $\approx$0.011         &$\approx$-0.41$\times10^{-4}$ &  0.5     &$\approx$-0.76$\times10^{-13}$ &$\approx$0.12    \\
SAX J1748.9-2021 & $1.81^{+0.25}_{-0.37}$ & $11.7^{+1.7}_{-1.7}$ &   $\approx$0.014         &$\approx$-0.5$\times10^{-4}$ &  0.5     &$\approx$-0.59$\times10^{-13}$ &$\approx$0.11    \\
EXO 1745-268 & $1.65^{+0.21}_{-0.31}$ & $10.5^{+1.6}_{-1.6}$ &    $\approx$0.017         &$\approx$-0.75$\times10^{-4}$ &  0.5     &$\approx$-0.59$\times10^{-13}$ &$\approx$0.14   \\
4U 1608-52 & $1.57^{+0.30}_{-0.29}$ & $9.8^{+1.8}_{-1.8}$ &   $\approx$0.02         &$\approx$-0.99$\times10^{-4}$ &  0.5     &$\approx$-0.33$\times10^{-12}$ &$\approx$0.17 \\
KS 1731-260 & $1.61^{+0.35}_{-0.37}$ & $10.0^{+2.2}_{-2.2}$ &  $\approx$0.018         &$\approx$-0.8$\times10^{-4}$ &  0.5     &$\approx$-0.27$\times10^{-12}$ &$\approx$0.13  \\
&&&&&&&\\
\end{tabular*}
\end{ruledtabular}
\end{table*}
\begin{table*}[t!]
\caption{\label{Table2}%
Values of physical quantities}
\begin{ruledtabular}
\begin{tabular*}{\textwidth}{lcccccccccc}
{{Pulsar}}                              &{$\rho|_{_{_{0}}}$} &      {$\rho|_{_{_{R}}}$} &   {$\frac{dp_r}{d\rho}|_{_{_{0}}}$}  &    {$\frac{dp_r}{d\rho}|_{_{_{R}}}$}
  & {$\frac{dp_t}{d\rho}|_{_{_{0}}}$} & {$\frac{dp_t}{d\rho}|_{_{_{R}}}$}&{$(\rho-p_r-2p_t)|_{_{_{0}}}$}&{$(\rho-p_r-2p_t)|_{_{_{R}}}$}&{$z|_{_{_{R}}}$}&\\ \hline
&&&&&&&&&&\\
4U 1724-207 &0.15$\times10^{-2}$     & .36$\times10^{-12}$  &   .5   &   .5      & .56&2.13&.76$\times10^{-3}$&.63$\times10^{-3}$& .009  \\
4U 1820-30  & 0.13$\times10^{-2}$     &0 &  .5  &   .5        &.59&1.87 &.63$\times10^{-4}$&.6$\times10^{-4}$& .0077   \\
SAX J1748.9-2021 &0.16$\times10^{-2}$&1.18$\times10^{-13}$&.5  &.5 &.56&2.16 &.81$\times10^{-3}$&.66$\times10^{-3}$&.0092    \\
EXO 1745-268  & 0.20$\times10^{-2}$&1.33$\times10^{-8}$&.5  &.5 &.56&2.16 &.1$\times10^{-2}$&.8$\times10^{-3}$&.0092   \\
4U 1608-52 & .24$\times10^{-2}$     & 6.53$\times10^{-14}$    &.5      &.5  & .55&2.26 &.12$\times10^{-2}$&.92$\times10^{-3}$& .0096 \\
KS 1731-260 &0.22$\times10^{-2}$ &  2.2$\times10^{-13}$  & .5 & .5 &.55&2.25 &.11$\times10^{-2}$&.83$\times10^{-3}$&.0096    \\

\end{tabular*}
\end{ruledtabular}
\end{table*}

In addition to 4U 1820-30, a similar analysis can be developed for other pulsars. In Tables I and II, we report the results for other observed systems.

\section{Discussion and conclusions}\label{S5}
In this research, we studied anisotropic spherically symmetric spacetime in the frame of MGT. We obtained a system of three differential equations in six unknowns. To put this system in a closed form, we used the formula of the radial component of the metric potential given by Tolman \citep{PhysRev.55.364}, ], a linear form of the radial EoS, and the constraint given by the mimetic field and the metric potential $g_{rr}$ given by Eq. (\ref{trans2}).  The solution of this system involves five constants, three of them are fixed using the junction condition, matching the interior solution with the Schwarzschild exterior, as well as the vanishing of the radial pressure at the boundary. The fourth constant is determined from the radial EoS, leaving the fifth constant to be determined from the study of a real compact star. The main features of this study can be summarized as follows:\vspace{0.2cm}\\
$\star$ To demonstrate that our solution is compatible with a real compact star, we used the stellar $4U1820-30$, which has mass  $1.46^{+0.21}_{-0.21} M_{\odot}$, and radius  $11.1^{+1.8}_{-1.8}$km \citep{Roupas:2020mvs,Das:2021qaq}. The use of the mass and the radius fixed the two constants  $s_1=0.01049285312$ and $s_2=-.4062000611\times10^{-4}$. Furthermore, the use of the vanishing of the radial pressure on the boundary yields $s_5=.1195529130$. Finally, using the linear radial EoS through the assumption of $s_3=0.5$ we obtained the constant $s_4=.7572649088\times 10^{-13}$. \vspace{0.2cm}\\
 $\star$ As shown in Figure  \ref{Fig:1}, the metric potentials have no singularity either at the center of the star or at the boundary. Furthermore, in Figure   \ref{Fig:1} we show how the junction condition is made between the interior and exterior Schwarzschild solution.\vspace{0.2cm}\\
$\star$ Figure  \ref{Fig:2}, shows that the density, radial, and tangential pressures are positive and decrease toward the center of the star.\vspace{0.2cm}\\
$\star$ Figure  \ref{Fig:3}, shows that the anisotropic force has a negative sign, meaning that it is attractive since  $p_t-p_r<0$. Also, figure  \ref{Fig:3}, 3 shows that the gradient of the density, radial, and tangential pressures are negative; this is a necessary condition for any real star configuration. Finally,  Figure   \ref{Fig:3} shows that the radial and tangential  EoS's  are not constant.
\vspace{0.2cm}\\

$\star$ { The importance of energy conditions in the frame of modified gravitational theories were discussed in detail in \citep{Capozziello:2014bqa,Capozziello:2013vna}. In the present study, we tested the impact of extra force which comes from the mimetic field and get the matter content more realistic, as in the case of all these stars, the energy conditions, i.e.,  NEC, WEC, DEC, SEC were all well satisfied which support that the matter destitution is normal matter (not any kind of exotic matter). In figure  \ref{Fig:4}, 4 shows that our model satisfies all the energy conditions.} Furthermore, Figure   \ref{Fig:5} shows that the radial and tangential speeds are <1, as required for any realistic star. Also in figure  \ref{Fig:5} the mass and compactness of our model are positive, and the red-shift is less than five i.e.,  $Z<5$.\vspace{0.2cm}\\
$\star$ Figure  \ref{Fig:6}, shows that the resulting model is stable because its adiabatic index is greater than  $4/3$.\vspace{0.2cm}\\
$\star$ Finally, we extended our study to other stars, as shown in Tables I and II, which confirms that our model is verified for these compact stars.\vspace{0.2cm}\\
$\star$ {It is interesting to analyze the solutions presented in this study in the frame  of more compact objects like neutron stars, to make
contact with events like the $GW190814$. These solutions have been discussed in details  in \citep{Astashenok:2021peo} in the frame of $f(R)$ gravitational theory. In
our case, we must be more careful because our approach applies to inhomogeneous solutions. However, pulsars with
spin less than $3ms$ or even the product of the merging of two neutron stars if it is a neutron star, can initially be quite
inhomogeneous, thus during the ring-down, our solution could be relevant. We hope to discuss this topic in future work
since such a study would require the implementation of a numerical recipe appropriately tailored to our solutions.}
\newpage
\newcommand{\eqref}{}
\begin{center}
{\bf{Appendix A}}
\end{center}
\begin{center}
{\bf The form of tangential pressure and anisotropic force }
\end{center}
\begin{eqnarray*}
&& p_t (r) =\frac {1}{32\pi\, \left( 1+s_1{r}^{2}+s_2{r}^{4} \right) ^{3}}\Big[144\,s_4\,\pi\,{r}^{4}s_2+128\,s_4\,\pi\,{r}^{2}s_1+208
\,s_4\,\pi\,{r}^{8}{s_2}^{2}+64\,{s_4}^{2}{\pi}^{2}{r}^{18}{s_2}^{4}+
176\,{r}^{4}s_4\,\pi\,{s_1}^{2}+26\,{r}^{4}s_3\,s_1s_2\nonumber\\
&& +384\,{r}^{6}
s_4\,\pi\,s_1s_2+40\,s_3\,{r}^{2}s_2+8\,s_1{r}^{4}s_2+12\,s_3\,s_1+32\,
s_4\,\pi+5\,{s_2}^{2}{r}^{6}+3\,{r}^{2}{s_1}^{2}+6\,{r}^{6}s_3\,{s_2}^{
2}+8\,{r}^{2}s_3\,{s_1}^{2}+64\,{s_2}^{3}{r}^{14}s_4\,\pi\,s_1s_3\nonumber\\
&& +96
\,{s_2}^{2}{r}^{12}s_4\,\pi\,{s_1}^{2}s_3+64\,{r}^{10}s_4\,\pi\,s_2{s_1
}^{3}s_3+304\,{r}^{10}\pi\,s_4\,{s_2}^{2}s_1s_3+272\,{r}^{8}s_4
\,\pi\,s_2{s_1}^{2}s_3+288\,{r}^{6}s_4\,\pi\,s_1s_2s_3+2\,{s_2}^{4}{r}^{
14}s_3\nonumber\\
&& +{s_2}^{4}{r}^{14}{s_3}^{2}+2\,{r}^{6}{s_1}^{4}s_3+{r}^{6}{
s_1}^{4}{s_3}^{2}+14\,{r}^{6}s_2{s_1}^{2}+4\,{r}^{8}{s_1}^{3}s_2+16\,{r}^{8}{
s_2}^{2}s_1+10\,{r}^{4}{s_1}^{3}s_3+6\,{r}^{4}{s_1}^{3}{s_3}^{2}+64\,{r}
^{2}{\pi}^{2}{s_4}^{2}+9\,{r}^{2}{s_1}^{2}{s_3}^{2}\nonumber\\
&& +25\,{r}^{6}{s_2}^
{2}{s_3}^{2}+6\,{r}^{10}{s_2}^{2}{s_1}^{2}+10\,{r}^{10}{s_2}^{3}{s_3}^
{2}+16\,{r}^{10}{s_2}^{3}s_3+4\,{s_2}^{3}{r}^{12}s_1+256\,{\pi}^{2}{s_4
}^{2}{s_2}^{3}{r}^{16}s_1+16\,\pi\,s_4\,{s_2}^{4}{r}^{16}s_3\nonumber\\
&& +384\,{s_2}^{
2}{r}^{14}{s_4}^{2}{\pi}^{2}{s_1}^{2}+64\,{s_2}^{3}{r}^{14}s_4\,\pi\,s_1
+256\,s_2{r}^{12}{s_4}^{2}{\pi}^{2}{s_1}^{3}+96\,{s_2}^{2}{r}^{12}s_4\,
\pi\,{s_1}^{2}+768\,{s_2}^{2}{r}^{12}s_1{\pi}^{2}{s_4}^{2}+112\,{s_2}^{3}{r}
^{12}s_4\,\pi\,s_3\nonumber\\
&& +64\,{r}^{10}s_4\,\pi\,s_2{s_1}^{3}+768\,{r}^{10}
s_2{s_1}^{2}{\pi}^{2}{s_4}^{2}+320\,{r}^{10}\pi\,s_4\,{s_2}^{2}s_1+16\,{r}
^{8}s_4\,\pi\,{s_1}^{4}s_3+304\,{r}^{8}s_4\,\pi\,s_2{s_1}^{2}+768\,{r
}^{8}s_2s_1{\pi}^{2}{s_4}^{2}\nonumber\\
&& +176\,{r}^{8}s_4\,\pi\,{s_2}^{2}s_3+80\,
{r}^{6}\pi\,s_4\,{s_1}^{3}s_3+112\,{r}^{4}s_4\,\pi\,{s_1}^{2}s_3
+80\,{r}^{4}s_4\,\pi\,s_2s_3+48\,{r}^{2}s_4\,\pi\,s_1s_3+16\,\pi
\,s_4\,{s_2}^{4}{r}^{16}+256\,{s_2}^{3}{r}^{14}{s_4}^{2}{\pi}^{2}\nonumber\\
&& +8\,{
s_2}^{3}{r}^{12}s_1s_3+4\,{s_2}^{3}{r}^{12}s_1{s_3}^{2}+112\,{s_2}^{3}{r}^
{12}s_4\,\pi+64\,{r}^{10}{s_4}^{2}{\pi}^{2}{s_1}^{4}+12\,{r}^{10}{s_2}
^{2}{s_1}^{2}s_3+6\,{r}^{10}{s_2}^{2}{s_1}^{2}{s_3}^{2}+384\,{r}^{10}{
s_4}^{2}{\pi}^{2}{s_2}^{2}\nonumber\\
&& +16\,{r}^{8}s_4\,\pi\,{s_1}^{4}+8\,{r}^{8}{s_1
}^{3}s_3\,s_2+4\,{r}^{8}{s_1}^{3}s_2{s_3}^{2}+256\,{r}^{8}{s_4}^{2}{
\pi}^{2}{s_1}^{3}+42\,{r}^{8}{s_2}^{2}s_1s_3+26\,{r}^{8}{s_2}^{2}s_1{s_3}^
{2}+96\,{r}^{6}\pi\,s_4\,{s_1}^{3}+36\,{r}^{6}{s_1}^{2}s_3\,s_2\nonumber\\
&& +22\,{r}
^{6}{s_1}^{2}s_2{s_3}^{2}+384\,{r}^{6}{s_4}^{2}{\pi}^{2}{s_1}^{2}+256\,
{r}^{6}{s_4}^{2}{\pi}^{2}s_2+30\,{r}^{4}s_1s_2{s_3}^{2}+256\,{r}^{4}s_1{
\pi}^{2}{s_4}^{2}+{s_2}^{4}{r}^{14}+6\,{r}^{10}{s_2}^{3}+{r}^{6}{s_1}^{4}+
4\,{r}^{4}{s_1}^{3}\Big]\,,\nonumber\\
&&
\Delta=\frac {{r}^{2}}{32\pi\, \left( 1+s_1{r}^{2}+s_2{r}^{4} \right) ^{3
}}\Big[  3\,{s_1}^{2}+16\,{r}^{6}s_4\,\pi\,{s_1}^{4}
+8\,{r}^{6}{s_1}^{3}s_3\,s_2+4\,{r}^{6}{s_1}^{3}s_2{s_3}^{2}+256\,{r}^{6
}{s_4}^{2}{\pi}^{2}{s_1}^{3}+26\,{r}^{6}s_1{s_2}^{2}{s_3}^{2}\nonumber\\
&& +30\,{r}^{
6}s_1{s_2}^{2}s_3+64\,{r}^{4}s_4\,\pi\,{s_1}^{3}+22\,{r}^{4}{s_1}^{2}s_2{
s_3}^{2}+24\,{r}^{4}{s_1}^{2}s_3\,s_2+384\,{r}^{4}{s_4}^{2}{\pi}^{
2}{s_1}^{2}+256\,{r}^{4}{s_4}^{2}{\pi}^{2}s_2+80\,{r}^{2}s_4\,\pi\,{s_1}
^{2}+30\,{r}^{2}s_1s_2{s_3}^{2}\nonumber\\
&& -14\,{r}^{2}s_2s_3\,s_1+256\,{r}^{2}s_1{
s_4}^{2}{\pi}^{2}+48\,{r}^{2}s_4\,\pi\,s_2+64\,{s_4}^{2}{\pi}^{2}{
r}^{16}{s_2}^{4}+16\,s_4\,\pi\,{s_2}^{4}{r}^{14}+256\,{s_2}^{3}{r}^{12}{
s_4}^{2}{\pi}^{2}+8\,{s_2}^{3}{r}^{10}s_1s_3\nonumber\\
&& +4\,{s_2}^{3}{r}^{10}s_1{
s_3}^{2}+80\,{s_2}^{3}{r}^{10}s_4\,\pi+64\,{r}^{8}{s_4}^{2}{\pi}^
{2}{s_1}^{4}+12\,{r}^{8}{s_2}^{2}{s_1}^{2}s_3+6\,{r}^{8}{s_2}^{2}{s_1}^{2}{
s_3}^{2}+384\,{r}^{8}{s_4}^{2}{\pi}^{2}{s_2}^{2}+112\,s_4\,\pi\,{
r}^{6}{s_2}^{2}\nonumber\\
&& +48\,s_4\,\pi\,s_1s_3+5\,{s_2}^{2}{r}^{4}+8\,s_1{r}^{2}s_2+4
\,{r}^{6}s_2{s_1}^{3}+16\,{r}^{6}{s_2}^{2}s_1+2\,{r}^{4}{s_1}^{4}s_3+{r}^{4}{
s_1}^{4}{s_3}^{2}+14\,{r}^{4}s_2{s_1}^{2}-18\,{r}^{4}s_3\,{s_2}^{2}+25\,
{r}^{4}{s_2}^{2}{s_3}^{2}\nonumber\\
&& +6\,{r}^{2}{s_1}^{3}s_3+6\,{r}^{2}{s_1}^{3}{
s_3}^{2}+{s_2}^{4}{r}^{12}{s_3}^{2}+4\,{s_2}^{3}{r}^{10}s_1+2\,{s_2}^{4}
{r}^{12}s_3+12\,{r}^{8}{s_2}^{3}s_3+10\,{r}^{8}{s_2}^{3}{s_3}^{2}
+6\,{r}^{8}{s_2}^{2}{s_1}^{2}+9\,{s_1}^{2}{s_3}^{2}+{r}^{4}{s_1}^{4}\nonumber\\
&& +4\,{r}
^{2}{s_1}^{3}+{s_2}^{4}{r}^{12}+6\,{r}^{8}{s_2}^{3}+64\,{s_4}^{2}{\pi}^{2}
+20\,s_3\,s_2-8\,s_3\,{s_1}^{2}+32\,s_4\,\pi\,s_1+224\,{r}^{8}s_4
\,\pi\,{s_2}^{2}s_1+16\,{r}^{6}s_4\,\pi\,{s_1}^{4}s_3+208\,{r}^{6}s_4
\,\pi\,s_2{s_1}^{2}\nonumber\\
&& +768\,{r}^{6}s_1s_2{s_4}^{2}{\pi}^{2}+176\,{r}^{6}s_4\,
\pi\,{s_2}^{2}s_3+80\,{r}^{4}s_4\,\pi\,{s_1}^{3}s_3+192\,{r}^{4}
s_4\,\pi\,s_1s_2+112\,{r}^{2}s_4\,\pi\,{s_1}^{2}s_3+80\,{r}^{2}s_4
\,\pi\,s_2s_3\nonumber\\
&& +256\,{s_4}^{2}{\pi}^{2}{s_2}^{3}{r}^{14}s_1+16\,s_4\,
\pi\,{s_2}^{4}{r}^{14}s_3+384\,{s_2}^{2}{r}^{12}{s_4}^{2}{\pi}^{2}{s_1}
^{2}+64\,{s_2}^{3}{r}^{12}s_4\,\pi\,s_1+256\,s_2{r}^{10}{s_4}^{2}{\pi}^{
2}{s_1}^{3}+96\,{s_2}^{2}{r}^{10}s_4\,\pi\,{s_1}^{2}\nonumber\\
&& +768\,{s_2}^{2}{r}^{10}s_1
{s_4}^{2}{\pi}^{2}+112\,{s_2}^{3}{r}^{10}s_4\,\pi\,s_3+64\,{r}^{8
}s_4\,\pi\,s_2{s_1}^{3}+768\,{r}^{8}s_2{s_1}^{2}{s_4}^{2}{\pi}^{2} +64\,{s_2}
^{3}{r}^{12}s_4\,\pi\,s_1s_3+96\,{s_2}^{2}{r}^{10}s_4\,\pi\,{s_1}^{2}
s_3\nonumber\\
&&+64\,{r}^{8}s_4\,\pi\,s_2{s_1}^{3}s_3+304\,{r}^{8}s_4\,\pi\,{
s_2}^{2}s_1s_3+272\,{r}^{6}s_4\,\pi\,s_2{s_1}^{2}s_3+288\,{r}^{4}s_4
\,\pi\,s_1s_2s_3 \Big]\,.   \hspace*{5cm}(\eqref{\mathbf A})
\end{eqnarray*}
\begin{center}
\newpage
\renewcommand{\eqref}{}
{\bf{Appendix B}}
\end{center}
\begin{center}
{{\bf The gradient of the tangential pressure}}
\end{center}
\begin{eqnarray*}
 &&p'_t=\frac {r}{16\pi\, \left( 1+s_1{r}
^{2}+s_2{r}^{4} \right) ^{4}}\Big[ -28\,s_3\,{s_1}^{2}+40\,s_3\,s_2+64\,{s_4}^
{2}{\pi}^{2}+9\,{s_1}^{2}{s_3}^{2}+15\,{s_2}^{2}{r}^{4}+{s_2}^{4}{r}^{12}
+15\,{r}^{8}{s_2}^{3}-{r}^{4}{s_1}^{4}+2\,{r}^{2}{s_1}^{3} \nonumber\\
 &&+3072\,{s_4}^{2}
{\pi}^{2}{s_2}^{3}{r}^{14}s_1+128\,s_4\,\pi\,{s_2}^{4}{r}^{14}s_3+4224
\,{s_2}^{2}{r}^{12}{s_4}^{2}{\pi}^{2}{s_1}^{2}+464\,{s_2}^{3}{r}^{12}s_4
\,\pi\,s_1+2560\,s_2{r}^{10}{s_4}^{2}{\pi}^{2}{s_1}^{3}+608\,{s_2}^{2}{r}^{
10}s_4\,\pi\,{s_1}^{2} \nonumber\\
 &&+3840\,{s_2}^{2}{r}^{10}s_1{s_4}^{2}{\pi}^{2}+320
\,{s_2}^{3}{r}^{10}s_4\,\pi\,s_3+336\,{r}^{8}s_4\,\pi\,s_2{s_1}^{3}+
3456\,{r}^{8}s_2{s_1}^{2}{s_4}^{2}{\pi}^{2}+656\,{r}^{8}s_4\,\pi\,{s_2}^
{2}s_1+64\,{r}^{6}s_4\,\pi\,{s_1}^{4}s_3 \nonumber\\
 &&+512\,{r}^{6}s_4\,\pi\,s_2{s_1}
^{2}+2048\,{r}^{6}s_2s_1{s_4}^{2}{\pi}^{2}+384\,{r}^{6}s_4\,\pi\,{s_2}^{
2}s_3+128\,{r}^{4}s_4\,\pi\,{s_1}^{3}s_3+368\,{r}^{4}s_4\,\pi
\,s_1s_2+128\,{r}^{2}s_4\,\pi\,{s_1}^{2}s_3 \nonumber\\
 &&+160\,{r}^{2}s_4\,\pi\,s_2
s_3+1664\,{s_4}^{2}{\pi}^{2}{s_2}^{3}{r}^{16}{s_1}^{2}+896\,{s_4}^{
2}{\pi}^{2}{s_2}^{4}{r}^{18}s_1+32\,s_4\,\pi\,{s_2}^{5}{r}^{18}s_3+1536
\,{s_2}^{2}{r}^{14}{s_4}^{2}{\pi}^{2}{s_1}^{3}+256\,{s_2}^{3}{r}^{14}s_4
\,\pi\,{s_1}^{2} \nonumber\\
 &&+144\,{s_2}^{4}{r}^{16}s_4\,\pi\,s_1+704\,s_2{r}^{12}{s_4}
^{2}{\pi}^{2}{s_1}^{4}+224\,{s_2}^{2}{r}^{12}s_4\,\pi\,{s_1}^{3}+96\,{r}^{
10}s_4\,\pi\,s_2{s_1}^{4}+16\,{r}^{8}s_4\,\pi\,{s_1}^{5}s_3+48\,s_4
\,\pi\,s_1s_3+96\,s_4\,\pi\,{r}^{2}s_2 \nonumber\\
 &&+256\,s_4\,\pi\,{r}^{6}{s_2}^{2
}+832\,{s_4}^{2}{\pi}^{2}{r}^{16}{s_2}^{4}+128\,s_4\,\pi\,{s_2}^{4}{r}
^{14}+1408\,{s_2}^{3}{r}^{12}{s_4}^{2}{\pi}^{2}-4\,{s_2}^{3}{r}^{10}s_1
s_3-8\,{s_2}^{3}{r}^{10}s_1{s_3}^{2}+256\,{s_2}^{3}{r}^{10}s_4\,\pi \nonumber\\
 &&+
576\,{r}^{8}{s_4}^{2}{\pi}^{2}{s_1}^{4}-6\,{r}^{8}{s_2}^{2}{s_1}^{2}s_3
-10\,{r}^{8}{s_2}^{2}{s_1}^{2}{s_3}^{2}+1152\,{r}^{8}{s_4}^{2}{\pi}^{
2}{s_2}^{2}+64\,{r}^{6}s_4\,\pi\,{s_1}^{4}-8\,{r}^{6}{s_1}^{3}s_3\,s_2-8
\,{r}^{6}{s_1}^{3}s_2{s_3}^{2}\nonumber\\
 &&+
576\,{r}^{8}{s_4}^{2}{\pi}^{2}{s_1}^{4}+1024\,{r}^{6}{s_4}^{2}{\pi}^{2}{s_1}^{3}
-16\,{r}^{6}{s_2}^{2}s_1{s_3}^{2}+64\,{r}^{6}{s_2}^{2}s_1s_3+112\,{r}^{4
}s_4\,\pi\,{s_1}^{3}-9\,{r}^{4}{s_1}^{2}s_2{s_3}^{2}+42\,{r}^{4}{s_1}^{2}
s_3\,s_2\nonumber\\
 &&+896\,{r}^{4}{s_4}^{2}{\pi}^{2}{s_1}^{2}+448\,{r}^{4}{s_4}^
{2}{\pi}^{2}s_2+96\,{r}^{2}s_4\,\pi\,{s_1}^{2}+60\,{r}^{2}s_1s_2{s_3}^{2}
-100\,{r}^{2}s_2s_3\,s_1+384\,{r}^{2}s_1{s_4}^{2}{\pi}^{2}+192\,{s_4}
^{2}{\pi}^{2}{r}^{20}{s_2}^{5}\nonumber\\
 &&+32\,s_4\,\pi\,{s_2}^{5}{r}^{18}+12\,{s_2}^{
3}{r}^{12}{s_1}^{2}s_3+8\,{s_2}^{4}{r}^{14}s_1s_3+6\,{s_2}^{3}{r}^{12}{s_1
}^{2}{s_3}^{2}+4\,{s_2}^{4}{r}^{14}s_1{s_3}^{2}+128\,{r}^{10}{s_4}
^{2}{\pi}^{2}{s_1}^{5}+8\,{r}^{10}{s_2}^{2}{s_1}^{3}s_3\nonumber\\
 &&+4\,{r}^{10}{s_2}^{2
}{s_1}^{3}{s_3}^{2}+16\,{r}^{8}s_4\,\pi\,{s_1}^{5}+2\,{r}^{8}{s_1}^{4}
s_3\,s_2+{r}^{8}{s_1}^{4}s_2{s_3}^{2}+3\,{s_1}^{2}+32\,s_4\,\pi\,s_1+{s_2}
^{5}{r}^{16}+16\,s_1{r}^{2}s_2+32\,{r}^{6}{s_2}^{2}s_1-4\,{r}^{4}{s_1}^{4}s_3\nonumber\\
 &&
-3\,{r}^{4}{s_1}^{4}{s_3}^{2}+19\,{r}^{4}s_2{s_1}^{2}-182\,{r}^{4}s_3
\,{s_2}^{2}+75\,{r}^{4}{s_2}^{2}{s_3}^{2}+4\,{r}^{2}{s_1}^{3}s_3-6\,{r
}^{2}{s_1}^{3}{s_3}^{2}-2\,{s_2}^{4}{r}^{12}s_3-3\,{s_2}^{4}{r}^{12}{
s_3}^{2}+4\,{s_2}^{3}{r}^{10}s_1\nonumber\\
 &&+4\,{r}^{8}{s_2}^{2}{s_1}^{2}-25\,{r}^{8}{s_2
}^{3}{s_3}^{2}+62\,{r}^{8}{s_2}^{3}s_3+144\,s_4\,\pi\,{s_2}^{4}{r}
^{16}s_3\,s_1+96\,{r}^{10}s_4\,\pi\,{s_1}^{4}s_3\,s_2+256\,{s_2}^{3}{r
}^{14}s_4\,\pi\,{s_1}^{2}s_3\nonumber\\
 &&+224\,{s_2}^{2}{r}^{12}s_4\,\pi\,{s_1}^{3
}s_3+480\,{s_2}^{3}{r}^{12}s_4\,\pi\,s_1s_3+640\,{s_2}^{2}{r}^{10}
s_4\,\pi\,{s_1}^{2}s_3+352\,{r}^{8}s_4\,\pi\,s_2{s_1}^{3}s_3+832\,
{r}^{8}s_4\,\pi\,{s_2}^{2}s_1s_3\nonumber\\
 &&+640\,{r}^{6}s_4\,\pi\,s_2{s_1}^{2}
s_3+544\,{r}^{4}s_4\,\pi\,s_1s_2s_3+{r}^{8}{s_1}^{4}s_2+4\,{r}^{10}{s_1}
^{3}{s_2}^{2}+2\,{s_2}^{5}{r}^{16}s_3+{s_2}^{5}{r}^{16}{s_3}^{2}+6\,{s_2
}^{3}{r}^{12}{s_1}^{2}+4\,{s_2}^{4}{r}^{14}s_1 \Big] \, , \nonumber\\
 &&\hspace*{17cm}(\eqref{\mathbf B})
\end{eqnarray*}
\begin{center}
\renewcommand{\eqref}{}
{\bf{Appendix C}}
\end{center}
\begin{center}
{{\bf The  tangential sound speed}}
\end{center}
\begin{eqnarray*}
 &&v_r{}^2=\frac{dp_t}{d\rho}=-\frac {1}{4 \left( 1+s_1{r}^{2}+s_2{r}^{4}
 \right)  \left( 12\,{s_2}^{2}{r}^{4}+3\,{r}^{6}{s_2}^{2}s_1+{r}^{8}{s_2}^{3}+
13\,s_1{r}^{2}s_2+3\,{r}^{4}s_2{s_1}^{2}-5\,s_2+5\,{s_1}^{2}+{r}^{2}{s_1}^{3}
 \right) }\nonumber\\
 &&\times\Big[-28\,s_3\,{s_1}^{2}+40\,s_3\,s_2+64\,{s_4}^{2}{\pi}^
{2}+9\,{s_1}^{2}{s_3}^{2}+15\,{s_2}^{2}{r}^{4}+{s_2}^{4}{r}^{12}+15\,{r}^
{8}{s_2}^{3}-{r}^{4}{s_1}^{4}+2\,{r}^{2}{s_1}^{3}+3072\,{s_4}^{2}{\pi}^{2}
{s_2}^{3}{r}^{14}s_1\nonumber\\
 &&+128\,s_4\,\pi\,{s_2}^{4}{r}^{14}s_3+4224\,{s_2}^{2}{
r}^{12}{s_4}^{2}{\pi}^{2}{s_1}^{2}+464\,{s_2}^{3}{r}^{12}s_4\,\pi\,s_1+
2560\,s_2{r}^{10}{s_4}^{2}{\pi}^{2}{s_1}^{3}+608\,{s_2}^{2}{r}^{10}s_4\,
\pi\,{s_1}^{2}+3840\,{s_2}^{2}{r}^{10}s_1{s_4}^{2}{\pi}^{2}\nonumber\\
 &&+320\,{s_2}^{3}{r
}^{10}s_4\,\pi\,s_3+336\,{r}^{8}s_4\,\pi\,s_2{s_1}^{3}+3456\,{r}^{8
}s_2{s_1}^{2}{s_4}^{2}{\pi}^{2}+656\,{r}^{8}s_4\,\pi\,{s_2}^{2}s_1+64\,{r}
^{6}s_4\,\pi\,{s_1}^{4}s_3+512\,{r}^{6}s_4\,\pi\,s_2{s_1}^{2}\nonumber\\
 &&+2048\,{
r}^{6}s_2s_1{s_4}^{2}{\pi}^{2}+384\,{r}^{6}s_4\,\pi\,{s_2}^{2}s_3+128
\,{r}^{4}s_4\,\pi\,{s_1}^{3}s_3+368\,{r}^{4}s_4\,\pi\,s_1s_2+128\,{r}
^{2}s_4\,\pi\,{s_1}^{2}s_3+160\,{r}^{2}s_4\,\pi\,s_2s_3
\nonumber\\
 &&+1664\,{
s_4}^{2}{\pi}^{2}{s_2}^{3}{r}^{16}{s_1}^{2}+896\,{s_4}^{2}{\pi}^{2}{s_2}
^{4}{r}^{18}s_1+32\,s_4\,\pi\,{s_2}^{5}{r}^{18}s_3+1536\,{s_2}^{2}{r}^{
14}{s_4}^{2}{\pi}^{2}{s_1}^{3}+256\,{s_2}^{3}{r}^{14}s_4\,\pi\,{s_1}^{2}+144\,{s_2}^{4}{r}^{16}s_4\,\pi\,s_1\nonumber\\
 &&+704\,s_2{r}^{12}{s_4}^{2}{\pi}^{2}{
s_1}^{4}+224\,{s_2}^{2}{r}^{12}s_4\,\pi\,{s_1}^{3}+96\,{r}^{10}s_4\,\pi
\,s_2{s_1}^{4}+16\,{r}^{8}s_4\,\pi\,{s_1}^{5}s_3+48\,s_4\,\pi\,s_1
s_3+96\,s_4\,\pi\,{r}^{2}s_2+256\,s_4\,\pi\,{r}^{6}{s_2}^{2}\nonumber\\
 &&+832\,{
s_4}^{2}{\pi}^{2}{r}^{16}{s_2}^{4}+128\,s_4\,\pi\,{s_2}^{4}{r}^{14}+
1408\,{s_2}^{3}{r}^{12}{s_4}^{2}{\pi}^{2}-4\,{s_2}^{3}{r}^{10}s_1s_3-8
\,{s_2}^{3}{r}^{10}s_1{s_3}^{2}3+256\,{s_2}^{3}{r}^{10}s_4\,\pi+576\,{r}
^{8}{s_4}^{2}{\pi}^{2}{s_1}^{4}\nonumber\\
 &&-6\,{r}^{8}{s_2}^{2}{s_1}^{2}s_3-10\,{r}
^{8}{s_2}^{2}{s_1}^{2}{s_3}^{2}+1152\,{r}^{8}{s_4}^{2}{\pi}^{2}{s_2}^{2
}+64\,{r}^{6}s_4\,\pi\,{s_1}^{4}-8\,{r}^{6}{s_1}^{3}s_3\,s_2-8\,{r}^{6}
{s_1}^{3}s_2{s_3}^{2}+1024\,{r}^{6}{s_4}^{2}{\pi}^{2}{s_1}^{3}\nonumber\\
 &&-16\,{r}^
{6}{s_2}^{2}s_1{s_3}^{2}+64\,{r}^{6}{s_2}^{2}s_1s_3+112\,{r}^{4}s_4\,
\pi\,{s_1}^{3}-9\,{r}^{4}{s_1}^{2}s_2{s_3}^{2}+42\,{r}^{4}{s_1}^{2}s_3\,
s_2+896\,{r}^{4}{s_4}^{2}{\pi}^{2}{s_1}^{2}+448\,{r}^{4}{s_4}^{2}{\pi}
^{2}s_2\nonumber\\
 &&+96\,{r}^{2}s_4\,\pi\,{s_1}^{2}+60\,{r}^{2}s_1s_2{s_3}^{2}-100\,{r
}^{2}s_2s_3\,s_1+384\,{r}^{2}s_1{s_4}^{2}{\pi}^{2}+192\,{s_4}^{2}{\pi
}^{2}{r}^{20}{s_2}^{5}+32\,s_4\,\pi\,{s_2}^{5}{r}^{18}+12\,{s_2}^{3}{r}^{
12}{s_1}^{2}s_3\nonumber\\
&&+8\,{s_2}^{4}{r}^{14}s_1s_3+6\,{s_2}^{3}{r}^{12}{s_1}^{2}{
s_3}^{2}+4\,{s_2}^{4}{r}^{14}s_1{s_3}^{2}+128\,{r}^{10}{s_4}^{2}{
\pi}^{2}{s_1}^{5}+8\,{r}^{10}{s_2}^{2}{s_1}^{3}s_3+4\,{r}^{10}{s_2}^{2}{s_1}^
{3}{s_3}^{2}+16\,{r}^{8}s_4\,\pi\,{s_1}^{5}\nonumber\\
 &&+2\,{r}^{8}{s_1}^{4}s_3
\,s_2+{r}^{8}{s_1}^{4}s_2{s_3}^{2}+3\,{s_1}^{2}+32\,s_4\,\pi\,s_1+{s_2}^{5}{r
}^{16}+16\,s_1{r}^{2}s_2+32\,{r}^{6}{s_2}^{2}s_1-4\,{r}^{4}{s_1}^{4}s_3-3\,{r
}^{4}{s_1}^{4}{s_3}^{2}+19\,{r}^{4}s_2{s_1}^{2}\nonumber\\
 &&-182\,{r}^{4}s_3\,{s_2}^{
2}+75\,{r}^{4}{s_2}^{2}{s_3}^{2}+4\,{r}^{2}{s_1}^{3}s_3-6\,{r}^{2}{s_1
}^{3}{s_3}^{2}-2\,{s_2}^{4}{r}^{12}s_3-3\,{s_2}^{4}{r}^{12}{s_3}^
{2}+4\,{s_2}^{3}{r}^{10}s_1+4\,{r}^{8}{s_2}^{2}{s_1}^{2}-25\,{r}^{8}{s_2}^{3}{
s_3}^{2}\nonumber\\
 &&+62\,{r}^{8}{s_2}^{3}s_3+144\,s_4\,\pi\,{s_2}^{4}{r}^{16}
s_3\,s_1+96\,{r}^{10}s_4\,\pi\,{s_1}^{4}s_3\,s_2+256\,{s_2}^{3}{r}^{14
}s_4\,\pi\,{s_1}^{2}s_3+224\,{s_2}^{2}{r}^{12}s_4\,\pi\,{s_1}^{3}
s_3+480\,{s_2}^{3}{r}^{12}s_4\,\pi\,s_1s_3\nonumber\\
 &&+640\,{s_2}^{2}{r}^{10}
s_4\,\pi\,{s_1}^{2}s_3+352\,{r}^{8}s_4\,\pi\,s_2{s_1}^{3}s_3+832\,
{r}^{8}s_4\,\pi\,{s_2}^{2}s_1s_3+640\,{r}^{6}s_4\,\pi\,s_2{s_1}^{2}
s_3+544\,{r}^{4}s_4\,\pi\,s_1s_2s_3\nonumber\\
 &&+{r}^{8}{s_1}^{4}s_2+4\,{r}^{10}{s_1}
^{3}{s_2}^{2}+2\,{s_2}^{5}{r}^{16}s_3+{s_2}^{5}{r}^{16}{s_3}^{2}+6\,{s_2
}^{3}{r}^{12}{s_1}^{2}+4\,{s_2}^{4}{r}^{14}s_1\Big]\, . \hspace*{5cm}(\eqref{\mathbf C})
\end{eqnarray*}
\renewcommand{\eqref}{}
\begin{center}
{\bf{Appendix D}}
\end{center}
\begin{center}
{{\bf The  adiabatic tangential index}}
\end{center}
\newpage
\begin{eqnarray*}
 &&\Gamma_t=- \Bigg( -28\,s_3\,{s_1}^{2}+40\,s_3\,s_2+64\,{s_4}^{
2}{\pi}^{2}+9\,{s_1}^{2}{s_3}^{2}+15\,{s_2}^{2}{r}^{4}+{s_2}^{4}{r}^{12}+
15\,{r}^{8}{s_2}^{3}-{r}^{4}{s_1}^{4}+2\,{r}^{2}{s_1}^{3}+3072\,{s_4}^{2}{
\pi}^{2}{s_2}^{3}{r}^{14}s_1\nonumber\\
 &&+128\,s_4\,\pi\,{s_2}^{4}{r}^{14}s_3+4224\,
{s_2}^{2}{r}^{12}{s_4}^{2}{\pi}^{2}{s_1}^{2}+464\,{s_2}^{3}{r}^{12}s_4\,
\pi\,s_1+2560\,s_2{r}^{10}{s_4}^{2}{\pi}^{2}{s_1}^{3}+608\,{s_2}^{2}{r}^{10}
s_4\,\pi\,{s_1}^{2}+3840\,{s_2}^{2}{r}^{10}s_1{s_4}^{2}{\pi}^{2}\nonumber\\
 &&+320\,{s_2
}^{3}{r}^{10}s_4\,\pi\,s_3+336\,{r}^{8}s_4\,\pi\,s_2{s_1}^{3}+3456
\,{r}^{8}s_2{s_1}^{2}{s_4}^{2}{\pi}^{2}+656\,{r}^{8}s_4\,\pi\,{s_2}^{2}s_1
+64\,{r}^{6}s_4\,\pi\,{s_1}^{4}s_3+512\,{r}^{6}s_4\,\pi\,s_2{s_1}^{2}\nonumber\\
 &&
+2048\,{r}^{6}s_2s_1{s_4}^{2}{\pi}^{2}+384\,{r}^{6}s_4\,\pi\,{s_2}^{2}
s_3+128\,{r}^{4}s_4\,\pi\,{s_1}^{3}s_3+368\,{r}^{4}s_4\,\pi\,s_1
s_2+128\,{r}^{2}s_4\,\pi\,{s_1}^{2}s_3+160\,{r}^{2}s_4\,\pi\,s_2
s_3\nonumber\\
 &&+1664\,{s_4}^{2}{\pi}^{2}{s_2}^{3}{r}^{16}{s_1}^{2}+896\,{s_4}^{
2}{\pi}^{2}{s_2}^{4}{r}^{18}s_1+32\,s_4\,\pi\,{s_2}^{5}{r}^{18}s_3+1536
\,{s_2}^{2}{r}^{14}{s_4}^{2}{\pi}^{2}{s_1}^{3}+256\,{s_2}^{3}{r}^{14}s_4
\,\pi\,{s_1}^{2}+144\,{s_2}^{4}{r}^{16}s_4\,\pi\,s_1\nonumber\\
 &&+704\,s_2{r}^{12}{s_4}
^{2}{\pi}^{2}{s_1}^{4}+224\,{s_2}^{2}{r}^{12}s_4\,\pi\,{s_1}^{3}+96\,{r}^{
10}s_4\,\pi\,s_2{s_1}^{4}+16\,{r}^{8}s_4\,\pi\,{s_1}^{5}s_3+48\,s_4
\,\pi\,s_1s_3+96\,s_4\,\pi\,{r}^{2}s_2+256\,s_4\,\pi\,{r}^{6}{s_2}^{2
}\nonumber\\
 &&+832\,{s_4}^{2}{\pi}^{2}{r}^{16}{s_2}^{4}+128\,s_4\,\pi\,{s_2}^{4}{r}
^{14}+1408\,{s_2}^{3}{r}^{12}{s_4}^{2}{\pi}^{2}-4\,{s_2}^{3}{r}^{10}s_1
s_3-8\,{s_2}^{3}{r}^{10}s_1{s_3}^{2}+256\,{s_2}^{3}{r}^{10}s_4\,\pi+
576\,{r}^{8}{s_4}^{2}{\pi}^{2}{s_1}^{4}\nonumber\\
 &&-6\,{r}^{8}{s_2}^{2}{s_1}^{2}s_3
-10\,{r}^{8}{s_2}^{2}{s_1}^{2}{s_3}^{2}+1152\,{r}^{8}{s_4}^{2}{\pi}^{
2}{s_2}^{2}+64\,{r}^{6}s_4\,\pi\,{s_1}^{4}-8\,{r}^{6}{s_1}^{3}s_3\,s_2-8
\,{r}^{6}{s_1}^{3}s_2{s_3}^{2}+1024\,{r}^{6}{s_4}^{2}{\pi}^{2}{s_1}^{3}\nonumber\\
 &&
-16\,{r}^{6}{s_2}^{2}s_1{s_3}^{2}+64\,{r}^{6}{s_2}^{2}s_1s_3+112\,{r}^{4
}s_4\,\pi\,{s_1}^{3}-9\,{r}^{4}{s_1}^{2}s_2{s_3}^{2}+42\,{r}^{4}{s_1}^{2}
s_3\,s_2+896\,{r}^{4}{s_4}^{2}{\pi}^{2}{s_1}^{2}+448\,{r}^{4}{s_4}^
{2}{\pi}^{2}s_2\nonumber\\
 &&+96\,{r}^{2}s_4\,\pi\,{s_1}^{2}+60\,{r}^{2}s_1s_2{s_3}^{2}
-100\,{r}^{2}s_2s_3\,s_1+384\,{r}^{2}s_1{s_4}^{2}{\pi}^{2}+192\,{s_4}
^{2}{\pi}^{2}{r}^{20}{s_2}^{5}+32\,s_4\,\pi\,{s_2}^{5}{r}^{18}+12\,{s_2}^{
3}{r}^{12}{s_1}^{2}s_3\nonumber\\
 &&+8\,{s_2}^{4}{r}^{14}s_1s_3+6\,{s_2}^{3}{r}^{12}{s_1
}^{2}{s_3}^{2}+4\,{s_2}^{4}{r}^{14}s_1{s_3}^{2}+128\,{r}^{10}{s_4}
^{2}{\pi}^{2}{s_1}^{5}+8\,{r}^{10}{s_2}^{2}{s_1}^{3}s_3+4\,{r}^{10}{s_2}^{2
}{s_1}^{3}{s_3}^{2}+16\,{r}^{8}s_4\,\pi\,{s_1}^{5}\nonumber\\
 &&+2\,{r}^{8}{s_1}^{4}
s_3\,s_2+{r}^{8}{s_1}^{4}s_2{s_3}^{2}+3\,{s_1}^{2}+32\,s_4\,\pi\,s_1+{s_2}
^{5}{r}^{16}+16\,s_1{r}^{2}s_2-2\,{s_2}^{4}{r}^{12}s_3-3\,{s_2}^{4}{r}^{12}
{s_3}^{2}+4\,{s_2}^{3}{r}^{10}s_1+4\,{r}^{8}{s_2}^{2}{s_1}^{2}\nonumber\\
 &&-25\,{r}^{8}{
s_2}^{3}{s_3}^{2}+62\,{r}^{8}{s_2}^{3}s_3+32\,{r}^{6}{s_2}^{2}s_1-4\,{r}
^{4}{s_1}^{4}s_3-3\,{r}^{4}{s_1}^{4}{s_3}^{2}+19\,{r}^{4}s_2{s_1}^{2}-
182\,{r}^{4}s_3\,{s_2}^{2}+75\,{r}^{4}{s_2}^{2}{s_3}^{2}+4\,{r}^{2}{
s_1}^{3}s_3\nonumber\\
 &&-6\,{r}^{2}{s_1}^{3}{s_3}^{2}+144\,s_4\,\pi\,{s_2}^{4}{r}
^{16}s_3\,s_1+96\,{r}^{10}s_4\,\pi\,{s_1}^{4}s_3\,s_2+256\,{s_2}^{3}{r
}^{14}s_4\,\pi\,{s_1}^{2}s_3+224\,{s_2}^{2}{r}^{12}s_4\,\pi\,{s_1}^{3
}s_3+480\,{s_2}^{3}{r}^{12}s_4\,\pi\,s_1s_3\nonumber\\
 &&+640\,{s_2}^{2}{r}^{10}
s_4\,\pi\,{s_1}^{2}s_3+352\,{r}^{8}s_4\,\pi\,s_2{s_1}^{3}s_3+832\,
{r}^{8}s_4\,\pi\,{s_2}^{2}s_1s_3+640\,{r}^{6}s_4\,\pi\,s_2{s_1}^{2}
s_3+544\,{r}^{4}s_4\,\pi\,s_1s_2s_3+4\,{s_2}^{4}{r}^{14}s_1\nonumber\\
 &&+{r}^{8}{s_1}
^{4}s_2+2\,{s_2}^{5}{r}^{16}s_3+{s_2}^{5}{r}^{16}{s_3}^{2}+6\,{s_2}^{3}{
r}^{12}{s_1}^{2}+4\,{r}^{10}{s_2}^{2}{s_1}^{3} \Bigg)\Bigg( 12\,s_1+20\,s_2{r
}^{2}+32\,s_4\,\pi+12\,s_3\,s_1+29\,{s_2}^{2}{r}^{6}+{s_2}^{4}{r}^{14}\nonumber\\
 &&+
10\,{r}^{10}{s_2}^{3}+{r}^{6}{s_1}^{4}+8\,{r}^{4}{s_1}^{3}+19\,{r}^{2}{s_1}^{2
}+256\,{s_4}^{2}{\pi}^{2}{s_2}^{3}{r}^{16}s_1+16\,s_4\,\pi\,{s_2}^{4}{r}
^{16}s_3+384\,{s_2}^{2}{r}^{14}{s_4}^{2}{\pi}^{2}{s_1}^{2}+64\,{s_2}^{3
}{r}^{14}s_4\,\pi\,s_1\nonumber\\
 &&+256\,s_2{r}^{12}{s_4}^{2}{\pi}^{2}{s_1}^{3}+96\,{
s_2}^{2}{r}^{12}s_4\,\pi\,{s_1}^{2}+768\,{s_2}^{2}{r}^{12}s_1{s_4}^{2}{\pi
}^{2}+112\,{s_2}^{3}{r}^{12}s_4\,\pi\,s_3+64\,{r}^{10}s_4\,\pi\,s_2
{s_1}^{3}+768\,{r}^{10}s_2{s_1}^{2}{s_4}^{2}{\pi}^{2}\nonumber\\
 &&+320\,{r}^{10}s_4\,
\pi\,{s_2}^{2}s_1+16\,{r}^{8}s_4\,\pi\,{s_1}^{4}s_3+304\,{r}^{8}s_4\,
\pi\,s_2{s_1}^{2}+768\,{r}^{8}s_2s_1{s_4}^{2}{\pi}^{2}+176\,{r}^{8}s_4\,
\pi\,{s_2}^{2}s_3+80\,{r}^{6}s_4\,\pi\,{s_1}^{3}s_3\nonumber\\
 &&+384\,{r}^{6}
s_4\,\pi\,s_1s_2+112\,{r}^{4}s_4\,\pi\,{s_1}^{2}s_3+80\,{r}^{4}s_4
\,\pi\,s_2s_3+48\,{r}^{2}s_4\,\pi\,s_1s_3+144\,s_4\,\pi\,{r}^{4}
s_2+128\,s_4\,\pi\,{r}^{2}s_1+208\,s_4\,\pi\,{r}^{8}{s_2}^{2}\nonumber\\
 &&+64\,{s_4
}^{2}{\pi}^{2}{r}^{18}{s_2}^{4}+16\,s_4\,\pi\,{s_2}^{4}{r}^{16}+256\,{s_2}
^{3}{r}^{14}{s_4}^{2}{\pi}^{2}+8\,{s_2}^{3}{r}^{12}s_1s_3+4\,{s_2}^{3}{
r}^{12}s_1{s_3}^{2}+112\,{s_2}^{3}{r}^{12}s_4\,\pi+64\,{r}^{10}{s_4
}^{2}{\pi}^{2}{s_1}^{4}\nonumber\\
 &&+12\,{r}^{10}{s_2}^{2}{s_1}^{2}s_3+6\,{r}^{10}{s_2}^
{2}{s_1}^{2}{s_3}^{2}+384\,{r}^{10}{s_4}^{2}{\pi}^{2}{s_2}^{2}+16\,{r
}^{8}s_4\,\pi\,{s_1}^{4}+8\,{r}^{8}{s_1}^{3}s_3\,s_2+4\,{r}^{8}{s_1}^{3}s_2
{s_3}^{2}+256\,{r}^{8}{s_4}^{2}{\pi}^{2}{s_1}^{3}\nonumber\\
 &&+26\,{r}^{8}{s_2}^{2
}s_1{s_3}^{2}+42\,{r}^{8}{s_2}^{2}s_1s_3+96\,{r}^{6}s_4\,\pi\,{s_1}^{3
}+22\,{r}^{6}{s_1}^{2}s_2{s_3}^{2}+36\,{r}^{6}{s_1}^{2}s_3\,s_2+384\,{r}
^{6}{s_4}^{2}{\pi}^{2}{s_1}^{2}+256\,{r}^{6}{s_4}^{2}{\pi}^{2}s_2\nonumber\\
 &&+176
\,{r}^{4}s_4\,\pi\,{s_1}^{2}+30\,{r}^{4}s_1s_2{s_3}^{2}+26\,{r}^{4}s_2
s_3\,s_1+256\,{r}^{4}s_1{s_4}^{2}{\pi}^{2}+40\,s_3\,{r}^{2}s_2+48\,s_1
{r}^{4}s_2+2\,{s_2}^{4}{r}^{14}s_3+{s_2}^{4}{r}^{14}{s_3}^{2}+4\,{s_2}^{
3}{r}^{12}s_1\nonumber\\
 &&+6\,{r}^{10}{s_2}^{2}{s_1}^{2}+10\,{r}^{10}{s_2}^{3}{s_3}^{2}+
16\,{r}^{10}{s_2}^{3}s_3+4\,{r}^{8}{s_1}^{3}s_2+28\,{r}^{8}{s_2}^{2}s_1+2\,{r
}^{6}{s_1}^{4}s_3+{r}^{6}{s_1}^{4}{s_3}^{2}+26\,{r}^{6}s_2{s_1}^{2}+6\,{
r}^{6}s_3\,{s_2}^{2}\nonumber\\
 &&+25\,{r}^{6}{s_2}^{2}{s_3}^{2}+10\,{r}^{4}{s_1}^{3
}s_3+6\,{r}^{4}{s_1}^{3}{s_3}^{2}+8\,{r}^{2}s_3\,{s_1}^{2}+9\,{r}
^{2}{s_1}^{2}{s_3}^{2}+64\,{r}^{2}{s_4}^{2}{\pi}^{2}+64\,{s_2}^{3}{r}
^{14}s_4\,\pi\,s_1s_3+96\,{s_2}^{2}{r}^{12}s_4\,\pi\,{s_1}^{2}s_3\nonumber\\
 &&+
64\,{r}^{10}s_4\,\pi\,s_2{s_1}^{3}s_3+304\,{r}^{10}s_4\,\pi\,{s_2}^{2
}s_1s_3+272\,{r}^{8}s_4\,\pi\,s_2{s_1}^{2}s_3+288\,{r}^{6}s_4\,\pi
\,s_1s_2s_3 \Bigg) \Bigg\{4 \Bigg( 12\,{s_2}^{2}{r}^{4}+3\,{r}^{6}{s_2}^{2}s_1+{r}
^{8}{s_2}^{3}\nonumber\\
 &&+13\,s_1{r}^{2}s_2+3\,{r}^{4}s_2{s_1}^{2}-5\,s_2+5\,{s_1}^{2}+{r}^{2}{s_1
}^{3} \Bigg)  \Bigg( 32\,s_4\,\pi+12\,s_3\,s_1+5\,{s_2}^{2}{r}^{6}+{
s_2}^{4}{r}^{14}+6\,{r}^{10}{s_2}^{3}+{r}^{6}{s_1}^{4}+4\,{r}^{4}{s_1}^{3}+3\,
{r}^{2}{s_1}^{2}\nonumber\\
 &&+256\,{s_4}^{2}{\pi}^{2}{s_2}^{3}{r}^{16}s_1+16\,s_4\,
\pi\,{s_2}^{4}{r}^{16}s_3+384\,{s_2}^{2}{r}^{14}{s_4}^{2}{\pi}^{2}{s_1}
^{2}+64\,{s_2}^{3}{r}^{14}s_4\,\pi\,s_1+256\,s_2{r}^{12}{s_4}^{2}{\pi}^{
2}{s_1}^{3}+96\,{s_2}^{2}{r}^{12}s_4\,\pi\,{s_1}^{2}\nonumber\\
 &&+768\,{s_2}^{2}{r}^{12}s_1
{s_4}^{2}{\pi}^{2}+112\,{s_2}^{3}{r}^{12}s_4\,\pi\,s_3+64\,{r}^{
10}s_4\,\pi\,s_2{s_1}^{3}+768\,{r}^{10}s_2{s_1}^{2}{s_4}^{2}{\pi}^{2}+320
\,{r}^{10}s_4\,\pi\,{s_2}^{2}s_1+16\,{r}^{8}s_4\,\pi\,{s_1}^{4}s_3\nonumber\\
 &&+
304\,{r}^{8}s_4\,\pi\,s_2{s_1}^{2}+768\,{r}^{8}s_2s_1{s_4}^{2}{\pi}^{2}+
176\,{r}^{8}s_4\,\pi\,{s_2}^{2}s_3+80\,{r}^{6}s_4\,\pi\,{s_1}^{3}
s_3+384\,{r}^{6}s_4\,\pi\,s_1s_2+112\,{r}^{4}s_4\,\pi\,{s_1}^{2}
s_3+80\,{r}^{4}s_4\,\pi\,s_2s_3\nonumber\\
 &&+48\,{r}^{2}s_4\,\pi\,s_1s_3+
144\,s_4\,\pi\,{r}^{4}s_2+128\,s_4\,\pi\,{r}^{2}s_1+208\,s_4\,\pi\,{
r}^{8}{s_2}^{2}+64\,{s_4}^{2}{\pi}^{2}{r}^{18}{s_2}^{4}+16\,s_4\,\pi\,
{s_2}^{4}{r}^{16}+256\,{s_2}^{3}{r}^{14}{s_4}^{2}{\pi}^{2}\nonumber\\
 &&+8\,{s_2}^{3}{r}
^{12}s_1s_3+4\,{s_2}^{3}{r}^{12}s_1{s_3}^{2}+112\,{s_2}^{3}{r}^{12}s_4
\,\pi+64\,{r}^{10}{s_4}^{2}{\pi}^{2}{s_1}^{4}+12\,{r}^{10}{s_2}^{2}{s_1}^{
2}s_3+6\,{r}^{10}{s_2}^{2}{s_1}^{2}{s_3}^{2}+384\,{r}^{10}{s_4}^{2
}{\pi}^{2}{s_2}^{2}\nonumber\\
 &&+16\,{r}^{8}s_4\,\pi\,{s_1}^{4}+8\,{r}^{8}{s_1}^{3}
s_3\,s_2+4\,{r}^{8}{s_1}^{3}s_2{s_3}^{2}+256\,{r}^{8}{s_4}^{2}{\pi}^
{2}{s_1}^{3}+26\,{r}^{8}{s_2}^{2}s_1{s_3}^{2}+42\,{r}^{8}{s_2}^{2}s_1s_3+
96\,{r}^{6}s_4\,\pi\,{s_1}^{3}+22\,{r}^{6}{s_1}^{2}s_2{s_3}^{2}\nonumber\\
 &&+36\,{r}
^{6}{s_1}^{2}s_3\,s_2+384\,{r}^{6}{s_4}^{2}{\pi}^{2}{s_1}^{2}+256\,{r}^
{6}{s_4}^{2}{\pi}^{2}s_2+176\,{r}^{4}s_4\,\pi\,{s_1}^{2}+30\,{r}^{4}s_1s_2
{s_3}^{2}+26\,{r}^{4}s_2s_3\,s_1+256\,{r}^{4}s_1{s_4}^{2}{\pi}^{2}\nonumber\\
 &&+
40\,s_3\,{r}^{2}s_2+8\,s_1{r}^{4}s_2+2\,{s_2}^{4}{r}^{14}s_3+{s_2}^{4}{r}^
{14}{s_3}^{2}+4\,{s_2}^{3}{r}^{12}s_1+6\,{r}^{10}{s_2}^{2}{s_1}^{2}+10\,{r}
^{10}{s_2}^{3}{s_3}^{2}+16\,{r}^{10}{s_2}^{3}s_3+4\,{r}^{8}{s_1}^{3}s_2\nonumber\\
 &&+
16\,{r}^{8}{s_2}^{2}s_1+2\,{r}^{6}{s_1}^{4}s_3+{r}^{6}{s_1}^{4}{s_3}^{2}
+14\,{r}^{6}s_2{s_1}^{2}+6\,{r}^{6}s_3\,{s_2}^{2}+25\,{r}^{6}{s_2}^{2}{
s_3}^{2}+10\,{r}^{4}{s_1}^{3}s_3+6\,{r}^{4}{s_1}^{3}{s_3}^{2}+8\,
{r}^{2}s_3\,{s_1}^{2}\nonumber\\
 &&+9\,{r}^{2}{s_1}^{2}{s_3}^{2}+64\,{r}^{2}{s_4
}^{2}{\pi}^{2}+64\,{s_2}^{3}{r}^{14}s_4\,\pi\,s_1s_3+96\,{s_2}^{2}{r}^{
12}s_4\,\pi\,{s_1}^{2}s_3+64\,{r}^{10}s_4\,\pi\,s_2{s_1}^{3}s_3+
304\,{r}^{10}s_4\,\pi\,{s_2}^{2}s_1s_3\nonumber\\
 &&+272\,{r}^{8}s_4\,\pi\,s_2{s_1}^{
2}s_3+288\,{r}^{6}s_4\,\pi\,s_1s_2s_3 \Bigg)  \left( 1+s_1{r}^{2}+s_2
{r}^{4} \right) \Bigg\}^{-1}\,. \hspace*{7cm}(\eqref{\mathbf D})
\end{eqnarray*}
\centerline{\bf Data availability}
The data underlying this article will be shared on any reasonable request to the corresponding author.

\end{document}